\documentclass[11pt]{article}

\usepackage[final]{acl}

\usepackage{times}
\usepackage{latexsym}

\usepackage[T1]{fontenc}

\usepackage[utf8]{inputenc}

\usepackage{microtype}

\usepackage{inconsolata}

\usepackage{graphicx}
\usepackage{enumitem}
\usepackage{xspace}
\usepackage{adjustbox}
\usepackage{booktabs}
\usepackage{MnSymbol}
\usepackage{multirow}
\usepackage{makecell}
\usepackage[table,xcdraw]{xcolor}
\usepackage{colortbl}
\usepackage{algpseudocode}
\usepackage{algorithm}
\usepackage{caption}

\usepackage{amsfonts}
\usepackage{graphicx}
\usepackage{pifont}
\usepackage{xcolor}
\usepackage{fontawesome}
\usepackage{makecell}
\usepackage{caption}

\title{Hybrid-Vector Retrieval for Visually Rich Documents: Combining Single-Vector Efficiency and Multi-Vector Accuracy}

\author{
\textbf{Juyeon Kim}$^{1}$ \quad
\textbf{Geon Lee}$^{1}$ \quad
\textbf{Dongwon Choi}$^{1}$ \quad
\textbf{Taeuk Kim}$^{2}$\thanks{Co-corresponding authors.} \quad
\textbf{Kijung Shin}$^{1}$\footnotemark[1] \\
$^{1}$KAIST \quad $^{2}$Hanyang University \\
\texttt{\{juyeonkim, geonlee0325, cookie000215, kijungs\}@kaist.ac.kr}  \\ \texttt{kimtaeuk@hanyang.ac.kr} \\
}

\begin{document}

\newcommand{\method}{\textsc{HEAVEN}\xspace}
\newcommand{\benchmark}{\textsc{ViMDoc}\xspace}
\newcommand{\vs}{\text{VS-page}\xspace}
\newcommand{\vsps}{\text{VS-pages}\xspace}
\newcommand{\uarrow}{\small{~($\uparrow$)}}
\newcommand{\darrow}{\small{~($\downarrow$)}}
\newcommand{\cmt}[1]{\textcolor{red}{#1}}
\newcommand{\tbe}[1]{\textcolor{blue}{#1}}

\newcommand\red[1]{\textcolor{red}{#1}}
\newcommand\blue[1]{\textcolor{blue}{#1}}
\newcommand\green[1]{\textcolor{green}{#1}}
\newcommand\orange[1]{\textcolor{orange}{#1}}
\newcommand\brown[1]{\textcolor{brown}{#1}}
\newcommand\olive[1]{\textcolor{olive}{#1}}

\newcommand\sblue[1]{\small\textcolor{blue}{#1}}
\newcommand\sred[1]{\small\textcolor{red}{#1}}
\maketitle

\begin{abstract}
Retrieval over visually rich documents is essential for tasks such as legal discovery, scientific search, and enterprise knowledge management.
Existing approaches fall into two paradigms: \textit{single-vector retrieval}, which is efficient but coarse, and \textit{multi-vector retrieval}, which is accurate but computationally expensive.
To address this trade-off, we propose \method, a plug-and-play two-stage hybrid-vector framework.
In the first stage, \method efficiently retrieves candidate pages using a single-vector method over Visually-Summarized Pages (VS-Pages), which assemble representative visual layouts from multiple pages.
In the second stage, it reranks candidates with a multi-vector method while filtering query tokens by linguistic importance to reduce redundant computations.
To evaluate retrieval systems under realistic conditions, we also introduce \benchmark, a benchmark for \textit{visually rich}, \textit{multi-document}, and \textit{long-document} retrieval.
Across four benchmarks, \method attains 99.87\% of the Recall@1 performance of multi-vector models on average while reducing per-query computation by 99.82\%, achieving efficiency and accuracy.
Our code and datasets are available at: \url{https://github.com/juyeonnn/HEAVEN}

\end{abstract}

\section{Introduction}
\label{sec:intro}

Document retrieval aims to retrieve relevant document pages from a corpus for a given query, with broad applications including legal discovery, scientific literature search, and enterprise knowledge management.
With the rise of large language models (LLMs), it has gained renewed attention as a core component of Retrieval-Augmented Generation (RAG), which grounds model responses in retrieved evidence to enhance factual reliability.

While traditional document retrieval methods have primarily relied on text representations, many real-world pages contain visually complex elements, such as charts, tables, and figures, that are crucial for answering queries, motivating the task of visual document retrieval (VDR).
To process such content, optical character recognition (OCR) or complex layout parsing has been employed, which increases indexing time and complexity.
Recently, Large Vision-Language Models (LVLMs) have enabled directly encoding each page as an image to obtain visual embeddings, simplifying the retrieval pipeline and improving performance on visually rich documents~\cite{fayssecolpali}.

Modern VDR methods fall into two paradigms: single-vector and multi-vector retrieval.
\textit{Single-vector retrieval} encodes a query and a document page into single embeddings, enabling efficient similarity computation via a dot product~\cite{yuvisrag,ma2024unifying}.
In contrast, \textit{multi-vector retrieval} encodes them into multiple token- or patch-level embeddings and computes fine-grained interactions across all query–page vector pairs~\cite{fayssecolpali,xu2025llama,xiao2025metaembed}.

\begin{figure}[t]
\centering
\includegraphics[width=0.95\linewidth]{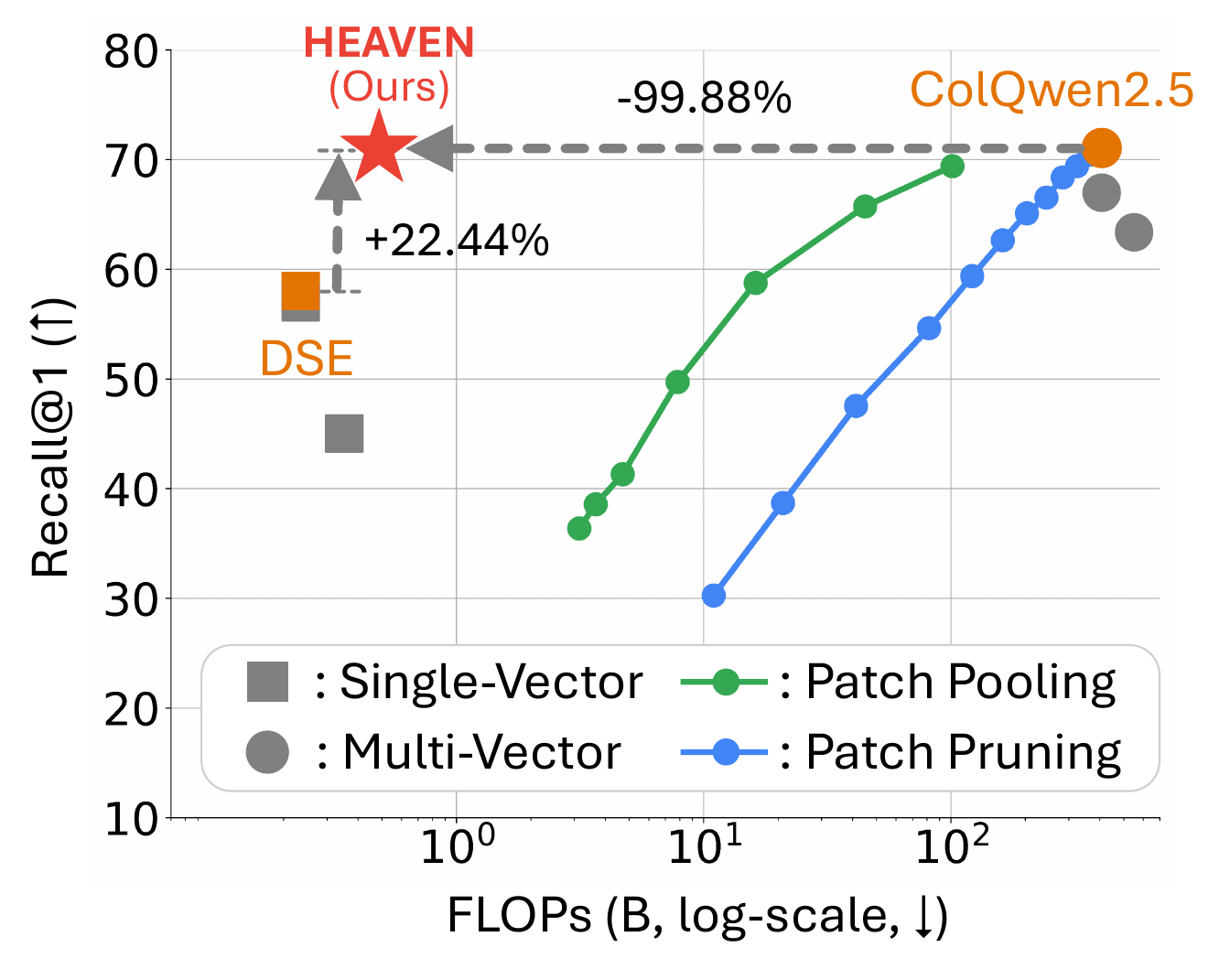}
\vspace{-5pt}
\caption{\label{fig:overview} Performance comparison of \method with (1) single-vector models, (2) multi-vector models, and (3) efficiency-oriented variants of multi-vector models (patch pooling and pruning). \method yields the best trade-off between efficiency and accuracy on the \benchmark benchmark. Refer to Section ~\ref{sec:exp:efficiency} for details.} 
\vspace{-15pt}
\end{figure}

Due to their design differences, single-vector and multi-vector retrieval methods exhibit a clear trade-off between efficiency and accuracy. 
While single-vector retrieval methods (e.g., DSE~\cite{ma2024unifying}) are highly efficient but less accurate,
multi-vector retrieval methods (e.g., ColQwen 2.5~\cite{fayssecolpali}) achieve higher accuracy at a substantially greater computational cost.

To address this efficiency-accuracy trade-off, we propose \method (\underline{\smash{\textbf{H}}}ybrid-vector retrieval for \underline{\smash{\textbf{E}}}fficient and \underline{\smash{\textbf{A}}}ccurate \underline{\smash{\textbf{V}}}isual multi-docum\underline{\smash{\textbf{EN}}}t), 
a hybrid framework that combines the efficiency of single-vector retrieval with the accuracy of multi-vector retrieval. 
Specifically, \method consists of two stages:
\begin{itemize}[leftmargin=*]
    \item \textbf{(Stage~1) Single-Vector Retrieval of Candidate Pages}: 
    Filtering is first performed at the level of our proposed \textit{visually-summarized pages}, which aggregate key visual elements across multiple pages, before applying filtering at the page level.
    \item \textbf{(Stage~2) Multi-Vector Reranking of Pages}: Candidate pages are reranked using only filtered query tokens, \textit{key tokens}, reducing computation while preserving accuracy. 
\end{itemize}
As shown in Figure~\ref{fig:overview}, \method provides a significantly improved efficiency–accuracy trade-off. Furthermore, \method is a plug-and-play framework. Its modular design allows the base encoder at each stage to be selected or swapped separately without further training, making it seamlessly adaptable to specific domains or easily upgraded to higher-performing models.

In addition, we present \benchmark (\underline{\smash{\textbf{Vi}}}sually-rich Long \underline{\smash{\textbf{M}}}ulti-\underline{\smash{\textbf{Doc}}}ument Retrieval Benchmark), a new benchmark for evaluating visual document retrieval under both multi-document and long-document settings.
Most existing VDR benchmarks either assume that queries can be resolved within a single document or focus on short documents. 
However, real-world applications
often require retrieval across massive, multi-document collections where individual documents often span dozens of pages. \benchmark addresses this gap.

Our contributions are summarized as follows:
\begin{itemize}[leftmargin=*]
\item \textbf{Method.} We introduce \method, a plug-and-play two-stage hybrid-vector retrieval framework that effectively addresses the efficiency-accuracy trade-off in visual document retrieval.

\item \textbf{Benchmark.} We propose \benchmark, a benchmark for visually rich, long-context, and multi-document retrieval. 

\item \textbf{Experiments.}
\method preserves 99.87\% of the state-of-the-art retrieval performance of the multi-vector model ColQwen2.5 (in terms of Recall@1) while reducing per-query FLOPs by 99.82\% across four multi-document benchmarks.
\end{itemize}

\section{Preliminaries}
\label{sec:prelim}
\begin{figure}[t]
    \centering
    \includegraphics[width=1\linewidth]{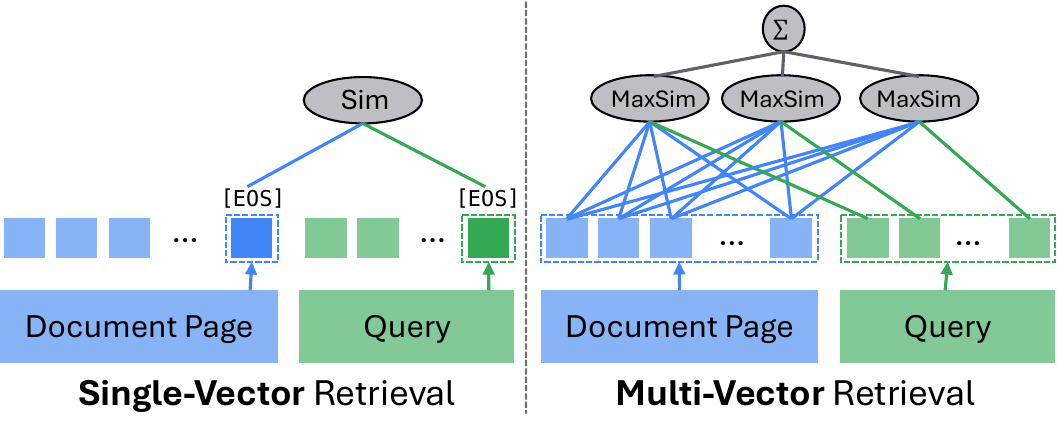}
    \vspace{-20pt}
    \caption{Single- and multi-vector retrieval models.}
    \label{fig:model-comparison} 
     \vspace{-5pt}
\end{figure}

We define the problem of document retrieval and review two main paradigms for addressing it.

\subsection{Problem Definition}
Let $\mathcal{D} = \{D_1, D_2, \ldots, D_{|\mathcal{D}|}\}$ denote a collection of documents.  
Each document $D_k \in \mathcal{D}$ is represented as an ordered sequence of pages, $D_k = ( P_{k,1}, P_{k,2}, \ldots, P_{k,|D_k|})$, where $P_{k,i}$ denotes the $i$-th page of the document $D_k$.
Let $\mathcal{P} = \bigcup_{D_k \in \mathcal{D}} D_k$ denote the set of all pages across the corpus.

We focus on \textit{page-level retrieval}.  
Given a query $q$, the goal is to find a set of ground-truth pages $\mathcal{P}_q \subseteq \mathcal{P}$, where the pages in $\mathcal{P}_q$ \textit{collectively} provide the information required to answer the query.

\subsection{Retrieval Frameworks}
A typical retrieval system represents a query $q$ and a page $P$ as embeddings in a shared latent space (see Figure~\ref{fig:model-comparison}).
Specifically, the query is represented as $\mathbf{E}_q \in \mathbb{R}^{n_q \times d}$ and the page as $\mathbf{E}_P \in \mathbb{R}^{n_P \times d}$, where $n_q$ and $n_P$ denote the number of vectors (e.g., tokens or patches) used to represent the query and page, respectively, and $d$ is the embedding dimension.
The system ranks pages by the relevance score using $\mathbf{E}_q$ and $\mathbf{E}_P$ and outputs the top-$K$ results.

\paragraph{Single-Vector Retrieval.}  
In single-vector retrieval, both queries and pages are represented by single embeddings rather than all token-level representations.  
Typically, this vector is obtained from a special token (e.g., an [EOS] token) or by aggregating over all token vectors (e.g., mean pooling).
The relevance score is computed as the similarity (e.g., dot product) between the query and page vectors:
\begin{equation*}
    S_\mathrm{SV}(q,P) = \langle \tilde{\mathbf{E}}_q, \tilde{\mathbf{E}}_P \rangle,
\end{equation*}
where $\tilde{\mathbf{E}}_q, \tilde{\mathbf{E}}_P \in \mathbb{R}^d$ denote the single-vector representations of the query $q$ and the page $P$, respectively.  
For each query, a single dot product is computed per page, resulting in $O(d \, |\mathcal{P}|)$ time to score all pages in the corpus.

\paragraph{Multi-Vector Retrieval.}
In multi-vector retrieval, the relevance score between a query and a page is computed by aggregating fine-grained interactions between all pairs of their embeddings:
\begin{equation*}
    S_\mathrm{MV}(q,P) = \sum_{i=1}^{n_q} \max_{j \in \{1,\cdots,n_P\}} \langle \mathbf{E}_{q}^{(i)}, \mathbf{E}_P^{(j)} \rangle,
\end{equation*}
where $\mathbf{E}_q^{(i)}$ and $\mathbf{E}_P^{(j)}$ denote the $i$-th and $j$-th row vectors of $\mathbf{E}_q$ and $\mathbf{E}_P$, respectively.  
Intuitively, this formulation finds, for each query token, the most relevant patch in the page and then sums these maximum similarities to produce the final score.
For a query $q$, scoring a page $P$ requires $O(d \, n_q \, n_P)$ time, and thus scoring all pages requires $O(d \, n_q \sum_{P \in \mathcal{P}} n_P)$ time.

\section{Analysis}
\label{sec:analysis}

\begin{figure}[t]
    \centering
    \vspace{-10pt}
    \includegraphics[width=1\linewidth]{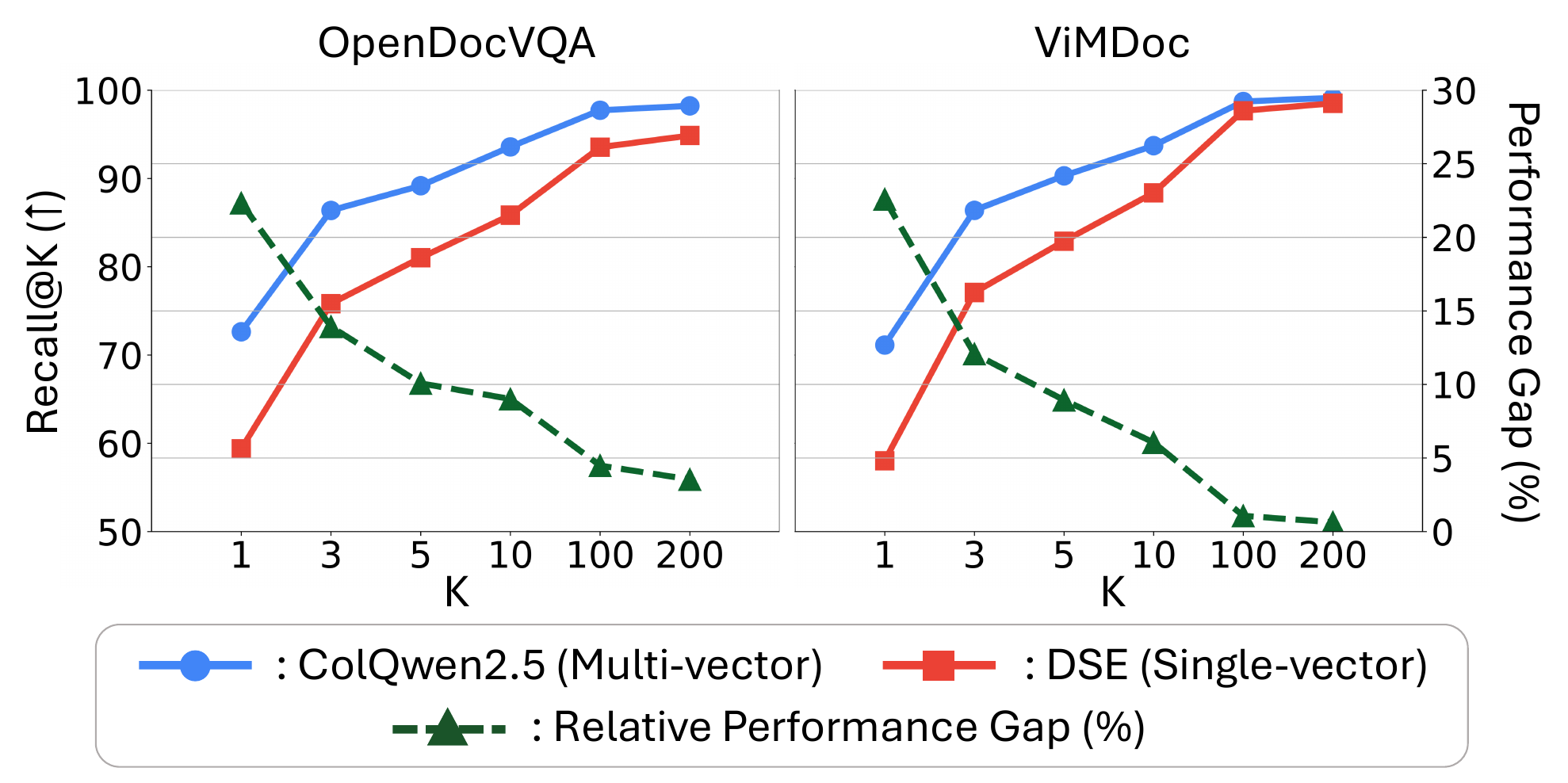}
    \caption{Multi-vector models (e.g., ColQwen2.5) outperform single-vector models (e.g., DSE), with pronounced gaps at fine-grained retrieval (K=1) but much smaller gaps at coarse-grained retrieval (K=200). }
    \label{fig:motivation}
\end{figure}

We empirically compare the two retrieval frameworks to examine their trade-offs.  
As shown in Figures~\ref{fig:overview} and ~\ref{fig:motivation}, we present two key observations:  

\paragraph{(Obs. 1) Single-Vector Retrieval is More Efficient Than Multi-Vector Retrieval.}  
Single-vector retrieval computes only one dot product per query-page pair, while multi-vector retrieval requires multiple comparisons. 
As shown in Figure~\ref{fig:overview}, single-vector retrieval is notably more efficient, requiring up to 99.94\% fewer FLOPs per query in \benchmark, making it scalable for large corpora.  

\paragraph{(Obs. 2) Single-Vector Retrieval can be Acceptable for Coarse-Grained Retrieval.}
As shown in Figure~\ref{fig:motivation}, single-vector retrieval is generally less accurate than multi-vector retrieval, as it cannot capture fine-grained token/patch–level interactions.  
However, this gap narrows as more candidates (larger top-$K$) are considered.  
For instance, the performance gap is 22.5\% at Recall@1 but only 0.63\% at Recall@200 in \benchmark.  
Thus, while multi-vector retrieval is superior for precise matching, single-vector retrieval can be sufficient for coarse-grained retrieval with broader candidate sets.

\section{Proposed Method: \method}
\label{sec:method}
\begin{figure}[t]
    \centering
    \includegraphics[width=1.0\linewidth]{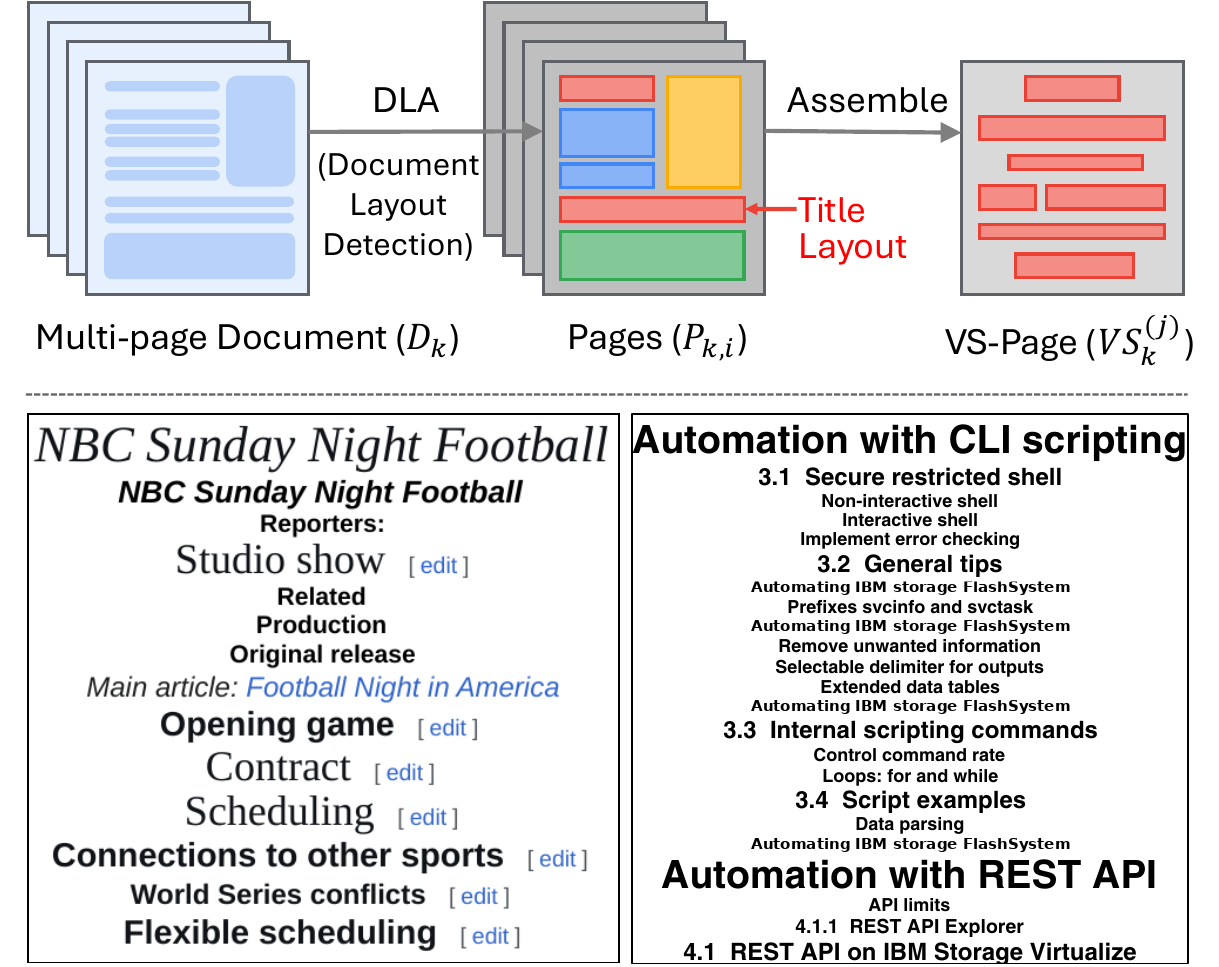}
    
    \caption{\textbf{Top:} Visually-summarized pages (\vsps) construction process. \textbf{Bottom:} Example outputs. More examples are provided in Appendix~\ref{app:vs-page-example}.}
    \label{fig:toc}
\end{figure}

\begin{figure*}[t]
    \centering
    
    \includegraphics[width=1.0\linewidth]{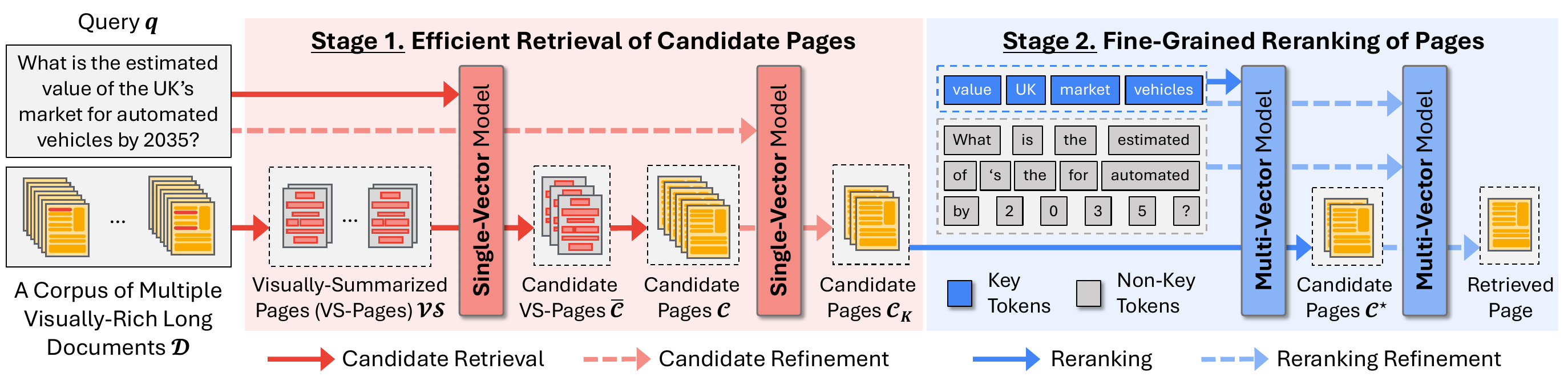}
    \vspace{-20pt}
    \caption{
    Overall pipeline of \method. 
    \textbf{(Stage~1)} Efficiently retrieves coarse candidate pages via single-vector retrieval, enhanced with visually-summarized pages (\vsps). 
    \textbf{(Stage~2)} Refines and reranks these candidates using a multi-vector model, enhanced with filtered key query tokens, for fine-grained retrieval.
    }
    \label{fig:main}
\end{figure*}

Motivated by our observations (Section~\ref{sec:analysis}), we propose \method (Figure~\ref{fig:main}), a two-stage hybrid-vector framework combining the \textit{efficiency} of single-vector and the \textit{accuracy} of multi-vector retrieval for visual document retrieval. Additionally, \method is plug-and-play. The single-vector and multi-vector models at each stage can be  selected or replaced without further training, making it seamlessly adaptable to specific domains or easily upgradeable to stronger models.

\subsection{Stage 1. Single-Vector Retrieval of Candidate Pages}
In the first stage, \method leverages our observation that single-vector retrieval is both efficient and sufficiently accurate for coarse-grained retrieval.
Thus, we use it to rapidly select a subset of candidate pages from a large document corpus.  

Despite its efficiency, existing single-vector retrieval models (e.g., DSE) compute similarities between the query and every page in the corpus, which scales linearly with the total number of pages, i.e., $|\mathcal{P}|$.
However, most pages are irrelevant to the query, making computation over the entire corpus redundant.  
Furthermore, many pages contain repetitive or uninformative content (e.g., recurring logos, headers, or boilerplate text), while only a few elements are actually informative and important for retrieving relevant pages to the query.

\paragraph{Visually-Summarized Pages (\vsps).}
To reduce this redundancy, we introduce \textit{visually-summarized pages} (\textit{\vsps}).
As illustrated in Figure~\ref{fig:toc}, multiple pages in a document are compressed into a smaller number of \vsps, each summarizing several pages by cropping representative layouts (specifically, title layouts) and assembling them into a single page.

\paragraph{\vs Construction (Indexing Phase Only).}
Given a document $D_k=(P_{k,1},P_{k,2},\ldots,P_{k,|D_k|})$, we first apply Document Layout Analysis ($\mathsf{DLA}$) to each page $P_{k,i}\in D_k$ to extract its title layouts:
\begin{equation*}
    T_{k,i} = \mathsf{DLA}(P_{k,i}) = \{t_{k,i}^{(1)},t_{k,i}^{(2)},\ldots,t_{k,i}^{(|T_{k,i}|)}\},
\end{equation*}
where $t_{k,i}^{(j)}$ is the $j$-th title layout extracted from page $P_{k,i}$.
Let $T_k = \bigcup_{i=1}^{|D_k|}T_{k,i}$ denote the set of layouts aggregated from all pages of document $D_k$.

Then, we partition $T_k$ into $\lceil |D_k|/r \rceil$ groups, where $r$ is a predefined reduction factor that controls the number of consecutive pages summarized by each \vs.
Each group $T_k^{(j)}$ thus contains approximately $|T_k|\cdot r/|D_k|$ title layouts.
The $j$-th \vs of $D_k$ is defined as:
\begin{equation*}
    \mathrm{VS}_k^{(j)} \;=\; \mathsf{Assemble}\big(T_k^{(j)}\big), \;\; j=1,\ldots,\lceil |D_k|/r \rceil,    
\end{equation*}
where $\mathsf{Assemble}(\cdot)$ is a function that composes the layouts into a single page.
Finally, the set of \vsps for $D_k$ is defined as:
\begin{equation*}
    \mathrm{VS}_k=\{\mathrm{VS}_k^{(1)},\ldots,\mathrm{VS}_k^{(\lceil |D_k|/r \rceil)}\}.    
\end{equation*}
Let $\mathcal{VS}=\bigcup_{D_k \in \mathcal{D}} \mathrm{VS}_k$ denote the set of all \vsps across the corpus.
Since each \vs summarizes multiple pages, $|\mathcal{VS}| < |\mathcal{P}|$ holds. We define $\Gamma(\mathrm{VS}_k^{(j)})=\{P_{k,i}\in\mathcal{P} : T_{k,i} \cap T_{k}^{(j)}\neq\emptyset\}$ as the set of pages associated with a \vs $\mathrm{VS}_k^{(j)}$.

Importantly, \vs construction is performed only once at indexing time, and thus does not add additional overhead during query-time inference. Further details are provided in Appendix~\ref{app:vs-page-detail}.

\paragraph{Candidate \vs Retrieval.}
Given a query $q$, we compute similarities with the constructed \vsps (fewer than the pages).
A single-vector retrieval model scores each $\mathrm{VS}\in\mathcal{VS}$ as:
\begin{equation*}
    S_\mathrm{SV}(q, \mathrm{VS}) = \langle \tilde{\mathbf{E}}_q, \tilde{\mathbf{E}}_\mathrm{VS} \rangle, \;\;\; \forall \, \mathrm{VS}\in\mathcal{VS},
\end{equation*}
where $\tilde{\mathbf{E}}_q, \tilde{\mathbf{E}}_{\mathrm{VS}} \in \mathbb{R}^d$ are the single-vector representation of the query and the \vs, respectively.
Then, we rank all \vsps based on their scores and retain the top-$(p_1\times 100)$\% as candidates, denoted by $\overline{\mathcal{C}}$, where $p_1$ is a hyperparameter.

\paragraph{Candidate Page Refinement.}
From the candidate \vsps $\overline{\mathcal{C}}\subset \mathcal{VS}$, we expand to their associated pages, i.e., $\mathcal{C} = \bigcup_{\mathrm{VS}\in \overline{\mathcal{C}}} \Gamma(\mathrm{VS})$. 
To refine these page candidates, we integrate both \vs-level and page-level scores.
For each candidate page $P\in \mathcal{C}$, the combined score is defined as:
\begin{equation*}
    S_\mathrm{SV}^\ast(q, P) = \alpha S_\mathrm{SV}(q, \Gamma^{-1}(P)) + (1-\alpha) S_\mathrm{SV}(q, P),
\end{equation*}
where $\Gamma^{-1}(P)$ denotes the \vs associated with page $P$, and $\alpha$ is a weighting hyperparameter. 
This score integrates both the \vs-level score $S_\mathrm{SV}(q, \Gamma^{-1}(P))$ and the page-level score $S_\mathrm{SV}(q,P)$. 
Then, we rank the pages in $\mathcal{C}$ by $S_\mathrm{SV}^\ast(q,P)$ and select the top-$K$ pages as the refined candidate set $\mathcal{C}_K$.

\subsection{Stage~2. Multi-Vector Reranking of Pages}
In the second stage, \method adopts a multi-vector framework to rerank the candidate set $\mathcal{C}_K$ obtained from Stage~1, computing fine-grained similarities across token- or patch-level embeddings of the query and each page for more precise ranking.

However, queries often include tokens that are less informative for the task (e.g., stopwords).
Multi-vector models nonetheless compute similarities between all query tokens and all page embeddings, causing redundant computation.

\begin{figure}[t]
    \centering
       \includegraphics[width=1\linewidth]{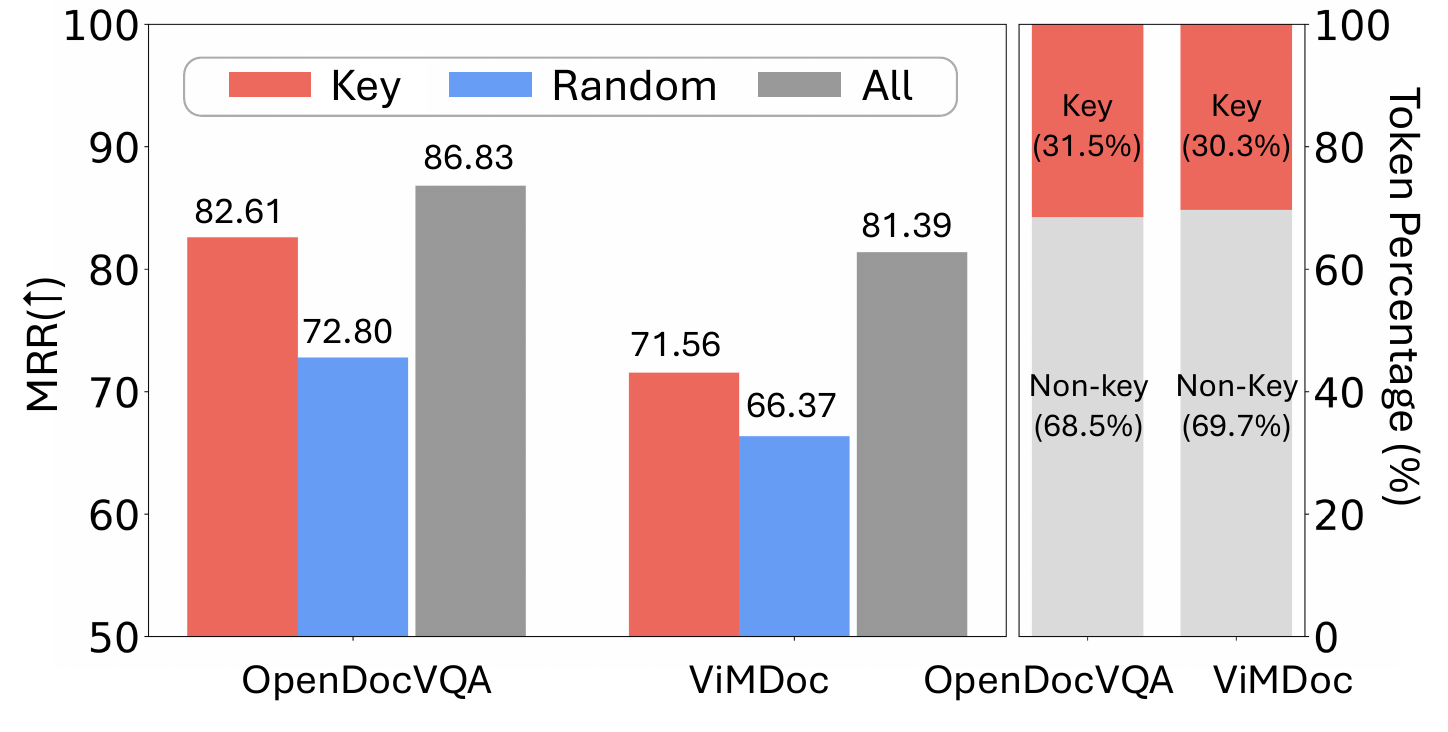}
    \vspace{-20pt}
    \caption{Using only \textit{key tokens} is reasonably effective, despite their small proportion ($\approx$ 30\%) within queries.}
    \label{fig:motivation-key}
\end{figure}

\paragraph{Reranking with Filtered Query Tokens.}
To reduce redundant computation, we filter query tokens based on their linguistic importance.  
Given a query $q$ with $n_q$ tokens $q=\{q^{(1)},q^{(2)},\ldots,q^{(n_q)}\}$, we apply Part-of-Speech (POS) tagging to identify a subset of \emph{key tokens} $q_\text{key} \subseteq q$ (e.g., nouns or named entities), which account for about 30\% of tokens on average, as described in Figure~\ref{fig:motivation-key}.  

The multi-vector relevance score between a filtered query $q_\text{key}$ and a candidate page $P\in \mathcal{C}_K$ is: 
\begin{equation*}
    S_\mathrm{MV}(q_\text{key}, P) = \sum_{i=1}^{|q_\text{key}|} \max_{j \in \{1,\ldots,n_P\}} 
    \langle \mathbf{E}_{q_\text{key}}^{(i)}, \mathbf{E}_P^{(j)} \rangle,
\end{equation*}
where $\mathbf{E}_{q_\text{key}}^{(i)}$ and $\mathbf{E}_P^{(j)}$ denote the embeddings of the $i$-th key token and the $j$-th page patch, respectively. 
This reduces the similarity computation by a factor of $|q_\text{key}|/n_q$.  
We use these scores to rerank the candidate pages $\mathcal{C}_K$ from Stage~1 and retain the top-$(p_2\times 100)\%$ as the final candidate set $\mathcal{C}^\star \subseteq \mathcal{C}_K$.  

Despite using fewer tokens, it achieves performance comparable to using all tokens and outperforms using the same number of random tokens, as shown in Figure~\ref{fig:motivation-key}.
While prior work reduces multi-vector retrieval complexity by pruning or pooling textual tokens~\cite{santhanam2022plaid, clavie2024reducing} or visual patches~\cite{fayssecolpali, ma2025towards, yan2025docpruner} of the documents, we demonstrate in Section~\ref{sec:exp} the effectiveness of filtering tokens of the queries instead.

\paragraph{Reranking Refinement.}
With the reduced set of candidate pages $\mathcal{C}^\star \subset \mathcal{P}$, we now perform a precise reranking using all query tokens, i.e., $S_\text{MV}(q,P)$.
For each candidate page $P \in \mathcal{C}^\star$, we refine the score by integrating the single-vector score $S_\text{SV}^\ast(q,P)$ from Stage~1 with the multi-vector score:
\begin{equation*}
S_\text{MV}^\ast(q,P) = \beta \, S_\text{SV}^\ast(q,P) + (1-\beta) \, S_\text{MV}(q,P),
\end{equation*}
where $\beta$ is a weighting hyperparameter.
Finally, we rank the pages in $\mathcal{C}^\star$ by $S_\text{MV}^\ast(q,P)$ to obtain the final retrieval results.

\section{Proposed Benchmark: \benchmark}
\label{sec:bench}

\begin{table}[t]

\setlength\tabcolsep{2.2pt} 
\setlength{\extrarowheight}{1pt}
\scalebox{0.63}{

\begin{tabular}{l|cc|cc|ccc}
\toprule
              \textbf{Benchmarks}            & \multicolumn{2}{c|}{\textbf{Document Page}} & \multicolumn{2}{c|}{\textbf{Query}} & \multirow{2}{*}{\textbf{Multi-}} & \multirow{2}{*}{\textbf{Long-}} & \multirow{2}{*}{\textbf{Cross-}} \\
\cmidrule(lr){2-3}
\cmidrule(lr){4-5}
\textbf{(Test Split)}       & \textbf{\#Total}    & \textbf{Avg./Doc}   & \textbf{\#Total}    & \textbf{\#Cross}    & \textbf{Doc}     & \textbf{Doc}     & \textbf{Query}      \\
\midrule
MP-DocVQA                 & 6.2k                 & 6.5        & 5.0k                 & -           &              &              &              \\
VisR-Bench                & 24.2k                 & 18.5       & 35.6k                 & -           &              &              &              \\

SlideVQA                  & 8.0k                 & 20.0       & 2.2k                 & 0.6k         &              &              & $\checkmark$ \\
\midrule

MMDocIR                   & 20.4k                 & 65.1       & 1.7k                  & 0.3k         &              & $\checkmark$ & $\checkmark$ \\
MMLongBench-Doc         & 6.5k                  & 47.5       & 1.1k                  & 0.4k         &              & $\checkmark$ & $\checkmark$ \\
LongDocURL                & 33.9k                 & 85.6       & 2.3k                  & 1.2k         &              & $\checkmark$ & $\checkmark$ \\
\midrule
ViDoRE                    & 8.3k                  & 1.0        & 3.8k                  & -           &            $\checkmark$ &              &              \\
ViDoSeek                  & 5.4k                  & 18.4       & 1.1k                  & -           & $\checkmark$ &              &              \\
REAL-MM-RAG               & 8.6k                  & 52.8       & 4.6k                  & -           &       $\checkmark$       & $\checkmark$ &              \\
M3DocVQA                   & 41.1k                 & 12.2       & 2.4k                  &    $\dagger$        & $\checkmark$ &              & $\checkmark$ \\
VisDoM                    & 21.0k                & 16.4       & 2.3k                  & $\dagger$           & $\checkmark$ &              & $\checkmark$ \\
OpenDocVQA                & 106.7k                & 3.1        & 2.5k                  & 0.3k         & $\checkmark$ &              & $\checkmark$ \\
\midrule
\rowcolor[HTML]{EFEFEF} 
\textbf{\benchmark(Ours)} & 76.3k                 & 55.4       & 10.9k                 & 0.7k         & $\checkmark$ & $\checkmark$ & $\checkmark$\\
\toprule
\end{tabular}}
\scriptsize{\textdagger  ~Page-level label not provided.} \\
\vspace{-5pt}

\caption{Comparison of VDR datasets.
\benchmark features multiple long documents with cross-page queries.
}
\label{tab:bench-comparison}

\end{table}

A corpus often contains numerous documents, each spanning multiple pages.
Thus, retrieval systems should be evaluated under two realistic settings: 
(i) the \textit{multi-document} setting, where relevant pages for a query may appear across documents, and
(ii) the \textit{long-document} setting, where individual documents are lengthy (e.g., $>$20 pages).
To address this, we propose \benchmark, a visual document retrieval benchmark that  jointly consider both settings.

\subsection{Existing Benchmarks}
Table~\ref{tab:bench-comparison} shows that existing benchmarks for VDR do not jointly consider both the multi-document and long-context settings.
Most focus on retrieving a relevant page from a single gold-standard document, which is often short (e.g., MP-DocVQA~\cite{tito2023hierarchical}, VisR-Bench~\cite{chen2025visr}, 
SlideVQA~\cite{tanaka2023slidevqa}) or, in some cases, long but still restricted to a single-document setting (e.g., MMLongBench-Doc~\cite{ma2024mmlongbench}, MMDocIR~\cite{dong2025mmdocir}, LongDocURL~\cite{deng2024longdocurl}).
Some benchmarks evaluate retrieval across multiple documents, but their documents are generally short, averaging only 1.0 - 18.4 pages (e.g., ViDoRe~\cite{fayssecolpali}, ViDoSeek~\cite{wang2025vidorag}, OpenDocVQA~\cite{tanaka2025vdocrag}, M3DocVQA~\cite{cho2024m3docrag}, VisDoM~\cite{suri2024visdom}).
As a result, these benchmarks fail to capture the combined challenge of retrieving relevant information across \textit{multiple long documents}.

\subsection{Design of \benchmark}
\benchmark is designed to evaluate VDR systems under both the \textit{multi-document} and \textit{long-document} settings.
It consists of visually rich documents and provides queries with page-level ground-truth annotations, including those with ground truth labels spanning multiple pages (i.e., cross-page queries).

\begin{table}[t]

\centering

\scalebox{0.75}{
\begin{tabular}{l|cc|c}
\toprule
& \multicolumn{2}{c|}{\textbf{Avg. Search Space}} & \multirow{2}{*}{\textbf{ Corpus}} \\
\cmidrule(lr){2-3}
\textbf{Benchmarks}  & \textbf{\# Page}     & \textbf{\# Doc}    &                 \\
\midrule
VisR-Bench             & 18.5                 & 1.0                  &     Single-Doc   \\
MMDocIR                & 65.1                 & 1.0                  &     Single-Doc     \\
MMLongBench-Doc        & 47.5                 & 1.0                  &     Single-Doc   \\
LongDocURL             & 85.6                 & 1.0                  &     Single-Doc  \\
REAL-MM-RAG            & 2151.0               & 40.8               &    Multi-Doc \\
\midrule
\rowcolor[HTML]{EFEFEF} 
\textbf{\benchmark(Ours)} & 76347.0     & 1379.0   &  Multi-Doc \\     
\bottomrule
\end{tabular}}
\caption{Comparison with sourced benchmarks. While most are limited to single-document 
settings with a small search space, \benchmark supports retrieval across a larger, multi-document corpus.}
\label{tab:bench-comparison-source}
\end{table}

\paragraph{Document Collection.}
We collect documents that (1) contain visually rich content such as figures, tables, and charts, and (2) consist of many pages, with an average length exceeding 20 pages. 
Specifically, we include documents from VisR-Bench~\cite{chen2025visr}, REAL-MM-RAG~\cite{wasserman2025real}, MMLongBench-Doc~\cite{ma2024mmlongbench}, MMDocIR~\cite{dong2025mmdocir}, and LongDocURL~\cite{deng2024longdocurl}, resulting in 1,379 documents spanning 76,347 pages. See Appendix~\ref{app:bench-collection} for details.

\paragraph{Query Processing.}
As shown in Table~\ref{tab:bench-comparison-source}, many sourced queries are designed 
for single-document settings. Some are therefore \textit{context-dependent}---relying on 
generic cues (e.g., ``what is the title'') or positional hints (e.g., ``from the last 
page'')---and thus unsuitable for multi-document retrieval. Following \citet{tanaka2025vdocrag} and \citet{wang2025vidorag}, we retain
only \textit{self-contained} queries with distinctive keywords such as named
entities or technical terms, enabling retrieval across the union of
all document pages. 45.8\% of queries identified as
\textit{context-dependent} are removed via a two-stage filtering pipeline: (1) heuristic
rule-based filtering and (2) LLM-based filtering. 
The retained queries can thus be evaluated in a multi-document setting with a 
larger search space than their sourced benchmarks.
Details are provided in Appendix~\ref{app:bench-query}.


\section{Experimental Results}
\label{sec:exp}

\begin{table*}[t]

 \vspace{-3pt}
\setlength\tabcolsep{1.7pt} 
\setlength{\extrarowheight}{1.75pt}
\scalebox{0.67}{
\begin{tabular}{cl|ccc|ccc|ccc|ccc|ccc}
\toprule
 & & \multicolumn{3}{c|}{\textbf{\benchmark} (Proposed)} & \multicolumn{3}{c|}{\textbf{OpenDocVQA}} & \multicolumn{3}{c|}{\textbf{ViDoSeek}} & \multicolumn{3}{c|}{\textbf{M3DocVQA}} & \multicolumn{3}{c}{\textbf{\textsc{AVERAGE     }}} \\
 \midrule
& \textbf{Model} & \textbf{R@1} & \textbf{R@3} & \textbf{FLOPs} & \textbf{R@1} & \textbf{R@3} & \textbf{FLOPs} & \textbf{R@1} & \textbf{R@3} & \textbf{FLOPs} & \textbf{R@1} & \textbf{R@3} & \textbf{FLOPs} & \textbf{R@1} & \textbf{R@3} & \textbf{FLOPs} \\
\midrule

\multirow{5}{*}{\rotatebox[origin=c]{90}{Single-Vector}} & BGE-M3 (dense) \textcolor{gray}{\textsuperscript{\faFileTextO}} & 44.89 & 61.62 & 0.156 & 38.67 & 52.45 & 0.165 & 54.82 & 76.01 & 0.011 & 57.56 & 77.11 & 0.084 & 48.99 & 66.80 & 0.104 \\ 
 & 
NV-Embed-V2 \textcolor{gray}{\textsuperscript{\faFileTextO}} & 50.06 & 69.08 & 0.625 & 47.37 & 64.59 & 0.658 & 66.46 & 83.89 & 0.044 & 62.65 & 85.77 & 0.336 & 56.64 & 75.83 & 0.416 \\ 

& VisRAG \textcolor{gray}{\textsuperscript{\faPictureO}} & 45.03 & 64.49 & 0.352 & 51.41 & 67.39 & 0.370 & 63.31 & 84.33 & 0.025 & 50.39 & 69.59 & 0.189 & 52.53 & 71.45 & 0.234 \\

& GME \textcolor{gray}{\textsuperscript{\faPictureO}} &57.01 & 76.62 & 0.235 & 54.19 & 71.93 & 0.247 & 71.19 & 89.40 & 0.017 & 59.85 & 77.83 & 0.126 & 60.56 & 78.95 & 0.156 \\

& DSE  \textcolor{gray}{\textsuperscript{\faPictureO}} & 58.03 & 77.08 & 0.235 & 59.38 & 75.82 & 0.247 & 69.53 & 87.13 & 0.017 & 55.14 & 71.30 & 0.126 & 60.52 & 77.83 & 0.156 \\
\midrule
\rowcolor[HTML]{EFEFEF} 
& \textbf{\method \small{(only Stage~1)}} & 57.64 & 76.58 & 0.134 & 59.89 & 76.05 & 0.147 & 69.35 & 87.39 & 0.010 & 59.40 & 73.96 & 0.026 & 61.57 & 78.50 & 0.079 \\

& \small{(vs. DSE)} & \small{\blue{99.34\%}} & \small{\blue{99.35\%}} & \small{\red{-42.76\%}} & \small{\blue{100.87\%}} & \small{\blue{100.32\%}} & \small{\red{-40.47\%}} & \small{\blue{99.75\%}} & \small{\blue{100.30\%}} & \small{\red{-42.40\%}} & \small{\blue{107.73\%}} & \small{\blue{103.73\%}} & \small{\red{-79.67\%}} & \small{\blue{101.74\%}} & \small{\blue{100.86\%}} & \small{\red{-49.31\%}}  \\
\midrule

\multirow{4}{*}{\rotatebox[origin=c]{90}{Multi-Vector}} &  BGE-M3 (multi)  \textcolor{gray}{\textsuperscript{\faFileTextO}} & 53.60 & 70.66 & 5863.387 & 46.60 & 62.80 & 888.097 & 56.74 & 76.01 & 30.777 & 56.88 & 76.96 & 1408.110 & 53.46 & 71.61 & 2047.590 \\ 

& ColPali \textcolor{gray}{\textsuperscript{\faPictureO}}     & 63.38 & 80.58 & 669.670 & 67.50 & 82.13 & 665.152 & 66.29 & 84.41 & 57.086 & 58.01 & 78.65 & 398.295 & 63.79 & 81.44 & 447.551 \\
& ColQwen2  \textcolor{gray}{\textsuperscript{\faPictureO}}   & 66.99 & 82.51 & 407.320 & 72.27 & 86.14 & 482.049 & 75.13 & 90.63 & 41.514 & 59.32 & 80.69 & 288.507 & 68.43 & 84.99 & 304.847 \\
& ColQwen2.5 \textcolor{gray}{\textsuperscript{\faPictureO}}  & 71.13 & 86.39 & 407.320 & 72.63 & 86.38 & 482.049 & 75.57 & 91.94 & 41.514 & 57.99 & 78.73 & 288.507 & 69.33 & 85.86 & 304.847 \\

\midrule
\rowcolor[HTML]{EFEFEF} 
& \textbf{\method} & 71.05 & 86.41 & 0.486 & 71.56 & 84.53 & 0.541 & 75.04 & 91.33 & 0.623 & 59.31 & 78.66 & 0.545 & 69.24 & 85.23 & 0.549 \\
& \small{(vs. ColQwen2.5)} & \small{\blue{99.88\%}} & \small{\blue{100.02\%}} & \small{\red{-99.88\%}} & \small{\blue{98.52\%}} & \small{\blue{97.86\%}} & \small{\red{-99.89\%}} & \small{\blue{99.30\%}} & \small{\blue{99.33\%}} & \small{\red{-98.50\%}} & \small{\blue{102.27\%}} & \small{\blue{99.90\%}} & \small{\red{-99.81\%}} & \small{\blue{99.87\%}} & \small{\blue{99.27\%}} & \small{\red{-99.82\%}} \\
\bottomrule
\end{tabular} 
}

\caption{
Efficiency–accuracy comparison with various single-vector and multi-vector models for visual document retrieval. We report Recall@\{1,3\} and per-query FLOPs (billions) for \method, compared with both single-vector and multi-vector models. The relative performance (\%) for Recall is highlighted in \blue{blue}, and the FLOPs is highlighted in \red{red}. \textcolor{gray}{\faFileTextO} indicates a textual embedding model and \textcolor{gray}{\faPictureO} indicates a visual embedding model. 
 }
\label{tab:main}
\end{table*}

In this section, we present the overall experimental setup and results, including comparisons with baselines, ablation study, efficiency analysis, hyperparameter analysis, and plug-and-play analysis.

\begin{figure*}[!ht]

\centering
\includegraphics[width=1\linewidth]{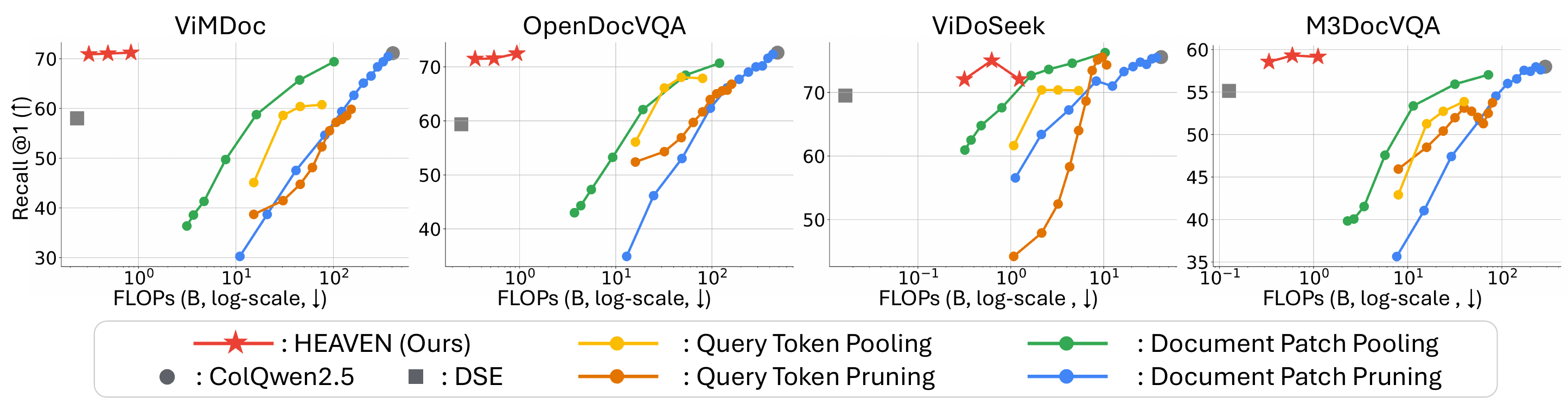}
\caption{Efficiency-accuracy comparison with four efficiency-oriented variants of the ColQwen2.5-based multi-vector model: (1) \textbf{document patch pooling}\tbe{~\cite{fayssecolpali}} and (2) \textbf{document patch pruning}~\cite{ma2025towards,yan2025docpruner} are applied to document patch embeddings; (3) \textbf{query token pooling} and (4) \textbf{query token pruning} are applied to query tokens using the same strategies.}
\label{fig:main-eff}
\end{figure*}

\subsection{Settings}

\paragraph{Models.} 
Two categories of retrieval models are utilized for evaluation. For single-vector retrieval, visual embedding models including VisRAG~\cite{yuvisrag}, GME~\cite{zhang2025bridging} and DSE ~\cite{ma2024unifying} are utilized, as well as textual embedding models NV-Embed-V2~\cite{lee2024nv} and BGE-M3 (dense)~\cite{chen2024bge}. For multi-vector retrieval, ColPali, ColQwen2, ColQwen2.5 ~\cite{fayssecolpali}, and BGE-M3 (multi-vec)~\cite{chen2024bge} are employed. Additional implementation details and model checkpoints are provided in Appendix~\ref{app:exp-detail-model}.

\paragraph{Datasets.} 
\method is tested on 4 benchmarks: \benchmark, OpenDocVQA~\cite{tanaka2025vdocrag}, ViDoSeek~\cite{wang2025vidorag} and M3DocVQA~\cite{cho2024m3docrag}. 
For OpenDocVQA, only the SlideVQA~\cite{tanaka2023slidevqa} and DUDE~\cite{van2023document} splits are used, as these are the only multi-page document splits. 
Detailed benchmark statistics are provided in Appendix~\ref{app:exp-detail-benchmark}.

\paragraph{Evaluation Metrics.} 

We evaluate retrieval performance using page-level Recall@\{1,3\}, except for M3DocVQA, where document-level metrics are used due to missing page-level labels.
Efficiency is measured by per-query FLOPs (billions) and latency (seconds).
See Appendix~\ref{app:exp-detail-metric} for details.

\paragraph{Implementation Details.}
\method uses DSE for Stage~1 and ColQwen2.5 for Stage~2.
For document layout analysis, it uses DocLayout-YOLO ~\cite{zhao2024doclayout}.
For each document $D_k$, the reduction factor $r$ is set to $\min(15, |D_k|)$.
Default hyperparameters are $\alpha=0.1$, $\beta=0.3$, $p_1=0.5$, $p_2=0.25$, and $K=200$.
For M3DocVQA, $\alpha$ and $\beta$ are set to 0.4 due to document-level evaluation.
\texttt{nltk} is used for POS tagging for \textit{key token} selection (Appendix~\ref{app:method-key-token}), and Tesseract~\cite{smith2007overview} is used to preprocess documents for textual retrieval.

\subsection{Main Results}

We first evaluate \method against state-of-the-art textual and visual retrieval models.
As shown in Table~\ref{tab:main}, on average, \method preserves 99.89\% of ColQwen2.5’s retrieval performance while reducing FLOPs by 99.82\%.
In M3DocVQA, its Stage~1 surpasses DSE by 7.73\% in Recall@1 with 79.67\% fewer FLOPs, and Stage~2 further improves performance by 2.27\% while using 99.81\% fewer FLOPs than ColQwen2.5.
Notably, the clear performance gap between textual and visual methods in \benchmark and OpenDocVQA reveals the unique challenges of visual document retrieval. 
Table~\ref{tab:cross-query} in Appendix~\ref{app:exp-extended} further reports 
results separately for cross-page and single-page query types.

\subsection{Comparison with Efficiency Variants}
\label{sec:exp:efficiency}

We evaluate \method's efficiency against four approaches designed to enhance efficiency in multi-vector retrieval models:
\textbf{(1) Document Patch Pooling}~\cite{fayssecolpali}, which pools adjacent patches within each page;
\textbf{(2) Document Patch Pruning}~\cite{ma2025towards,yan2025docpruner}, which randomly removes patch embeddings;
\textbf{(3) Query Token Pooling}, which aggregates adjacent special query tokens; and
\textbf{(4) Query Token Pruning}, which randomly drops them. 
Each approach is evaluated under varying pooling factors and pruning ratios.
Refer to Appendix~\ref{app:exp-detail-eff} for details.

Figure~\ref{fig:main-eff} shows that \method achieves the best efficiency-accuracy trade-off.
While pooling generally outperforms pruning, its performance drops as FLOPs decrease.
In contrast, \method attains higher accuracy with significantly fewer FLOPs. 

\subsection{Ablation Study}
To verify the effectiveness of each component in \method, we conduct ablation studies comparing it with its variants.
For Stage~1, we evaluate two variants, one that omits \vs construction and another that skips candidate refinement. 
Table~\ref{tab:ablation} reports that without \vsps, FLOPs notably increase, as all raw pages are compared, while yielding marginal accuracy gains.
Skipping candidate refinement leads to a severe performance drop. 

For Stage~2, we examine variants that disable query token filtering and reranking refinement. 
Without query filtering, FLOPs increase as similarities are computed for all query tokens, including redundant ones.
Without reranking refinement, performance degrades significantly. 
These results confirm the importance of the elements in two stages of \method for both efficiency and accuracy.

\begin{table}[t]

\setlength\tabcolsep{1.64pt} 
\setlength{\extrarowheight}{2pt}

\scalebox{0.69}{
\begin{tabular}{l|ccc|ccc}
\toprule
\multicolumn{1}{l}{} & \multicolumn{3}{|c}{\textbf{\benchmark} (Proposed)} & \multicolumn{3}{|c}{\textbf{OpenDocVQA}} \\

\midrule
\rowcolor[HTML]{EFEFEF} \textbf{Stage 1} & \textbf{R@100} & \textbf{R@200} & \textbf{FLOPs} & \textbf{R@100} & \textbf{R@200} & \textbf{FLOPs} \\
\midrule

\textbf{\method}  & 97.20          & 97.96          & 0.134          & 93.02               & 94.59          & 0.147          \\
\; w/o \vsps & 97.68          & 98.51          & 0.235          & 93.54               & 94.86          & 0.247          \\
\; w/o refinement & 59.47          & 68.02          & 0.017          & 59.47               & 68.02          & 0.024          \\
\midrule
\rowcolor[HTML]{EFEFEF} \textbf{Stage 2} & \textbf{R@1} & \textbf{R@3} & \textbf{FLOPs} & \textbf{R@1} & \textbf{R@3} & \textbf{FLOPs} \\

\midrule

\textbf{\method} & 71.05          & 86.41          & 0.486          & 71.56               & 84.53          & 0.541          \\
\; w/o query filtering & 71.08          & 86.38          & 0.871          & 71.76               & 85.10          & 0.957          \\
\; w/o refinement  & 58.05          & 77.08          & 0.358          & 59.53               & 75.82          & 0.402         \\
\bottomrule
\end{tabular}}
\caption{Ablation study of Stage~1 and Stage~2.
Each component of \method is effective for coarse-grained retrieval (Recall@\{100,200\}) in Stage 1 and fine-grained retrieval (Recall@\{1,3\}) in Stage 2.}
\label{tab:ablation}
\vspace{-5pt}
\end{table}

\subsection{Efficiency Analysis}

To evaluate the efficiency of \method, we measure two types of latency: 
\textbf{(1) Offline Indexing Latency}, referring to the time required to preprocess documents before retrieval, and 
\textbf{(2) Online Retrieval Latency}, referring to the time required to process a query during retrieval.\footnote{For fair comparison, all experiments are conducted on 16 Intel(R) Xeon(R) Platinum 8480C CPUs and a single NVIDIA H200 GPU. All results are averaged over five runs.}

As shown in Table~\ref{tab:latency}, \method requires additional offline processing time for \vs construction.
However, its online retrieval latency is \textit{significantly} lower than that of multi-vector methods, e.g., 99.88\% faster than ColQwen2.5 per query. 
Despite this efficiency, as shown in Table~\ref{tab:main}, \method achieves accuracy comparable to multi-vector models while substantially outperforming single-vector ones, confirming its strong trade-off between efficiency and accuracy. 

\begin{table}[t]

\begin{adjustbox}{max width=1.0\columnwidth}
\begin{tabular}{cl|ccc|cc}
\toprule
& & \multicolumn{5}{c}{\textbf{\benchmark} (Proposed)} \\
\cmidrule{3-7}
& & \multicolumn{3}{c|}{\textbf{Offline Indexing (min)}} & \multicolumn{2}{c}{\textbf{Online Retrieval (sec,~B)}} \\
\cmidrule{3-5}
\cmidrule{6-7}
& & OCR & \vs & Encode & Latency &  FLOPs \\
\midrule
\multirow{2}{*}{\rotatebox[origin=c]{90}{Single}} & 
NV-Embed-V2 \textcolor{gray}{\textsuperscript{\faFileTextO}} 
& 473.0& NA& 99.5 & 0.133 & 0.625 \\
& DSE \textcolor{gray}{\textsuperscript{\faPictureO}} 
& NA & NA & 146.7 & 0.115 & 0.235\\
\midrule
\multirow{2}{*}{\rotatebox[origin=c]{90}{Multi}} &
BGE-M3 (multi) \textcolor{gray}{\textsuperscript{\faFileTextO}} 
& 473.0 & NA & 32.3 & 20953.468 & 5863.387 \\
& ColQwen2.5 \textcolor{gray}
{\textsuperscript{\faPictureO}} 
& NA & NA & 129.1 &  2006.361 & 407.320 \\
\midrule
\rowcolor[HTML]{EFEFEF}
& \textbf{\method}
& NA & 58.9 & 286.4 & 2.412 & 0.486  \\
& \;\;Stage~1  
& NA & 58.9 &  157.3 & 0.079& 0.134 \\
& \;\;Stage~2  
& NA & NA & 129.1  & 2.333& 0.352 \\

\bottomrule
\end{tabular}
\end{adjustbox}
\caption{Efficiency analysis for offline indexing and online retrieval on \benchmark. Offline indexing efficiency is reported for the full document corpus, and online retrieval efficiency is reported per query. 
\vs construction latency includes $\mathsf{DLA}$ and $\mathsf{Assemble}$.}
\label{tab:latency}
\vspace{-5pt}

\end{table}

\subsection{Hyperparameter Analysis}

\begin{figure}[t]
    \centering
    \includegraphics[width=1\linewidth]{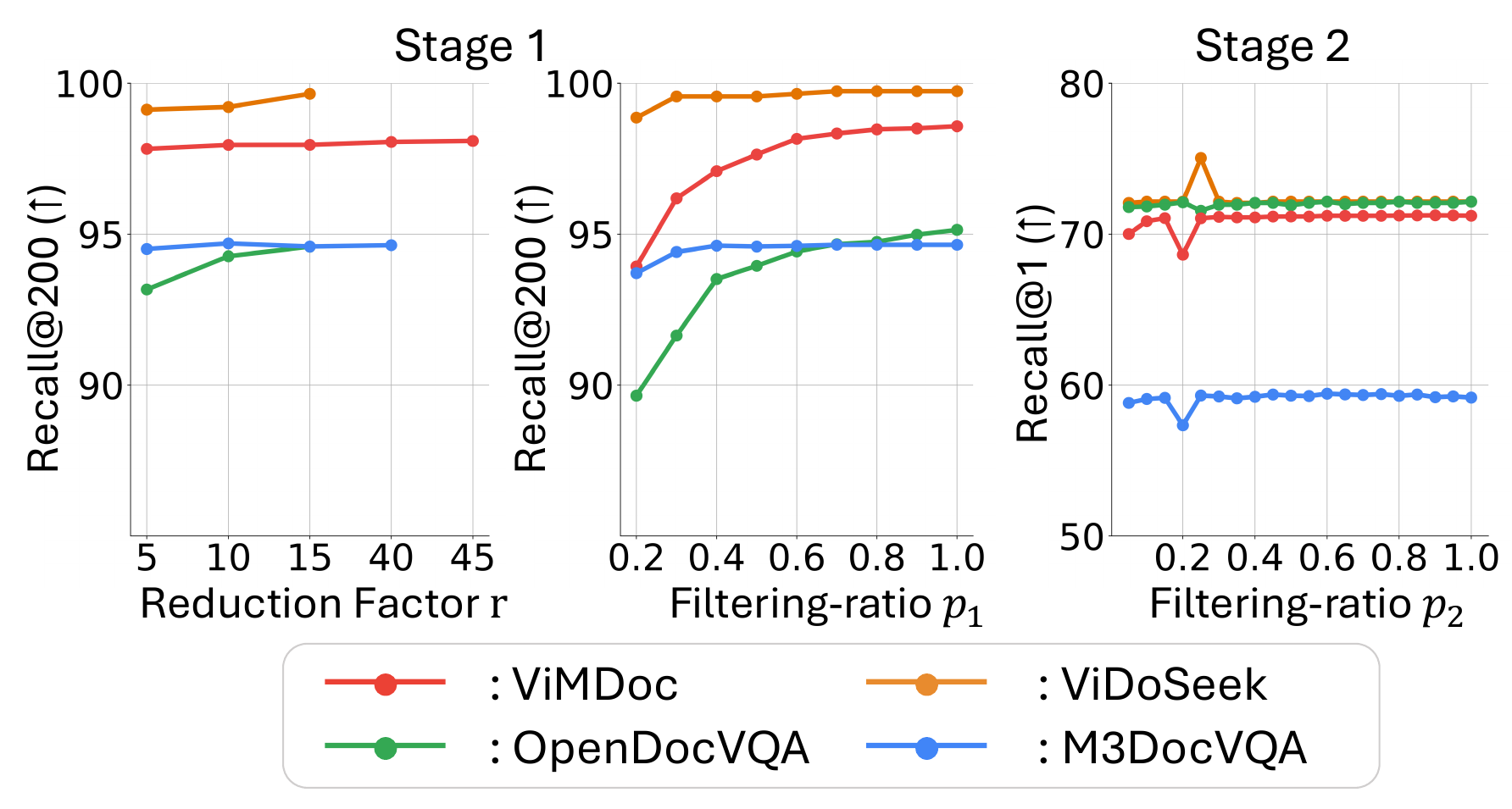}
    \vspace{-15pt}
    \caption{Hyperparameter analysis for reduction factor $r$, and filtering ratios $p_1$ and $p_2$.}
    \label{fig:hyper-anlysis}
    \vspace{-5pt}
\end{figure}

\begin{figure}[t]
    \centering
    \includegraphics[width=1\linewidth]{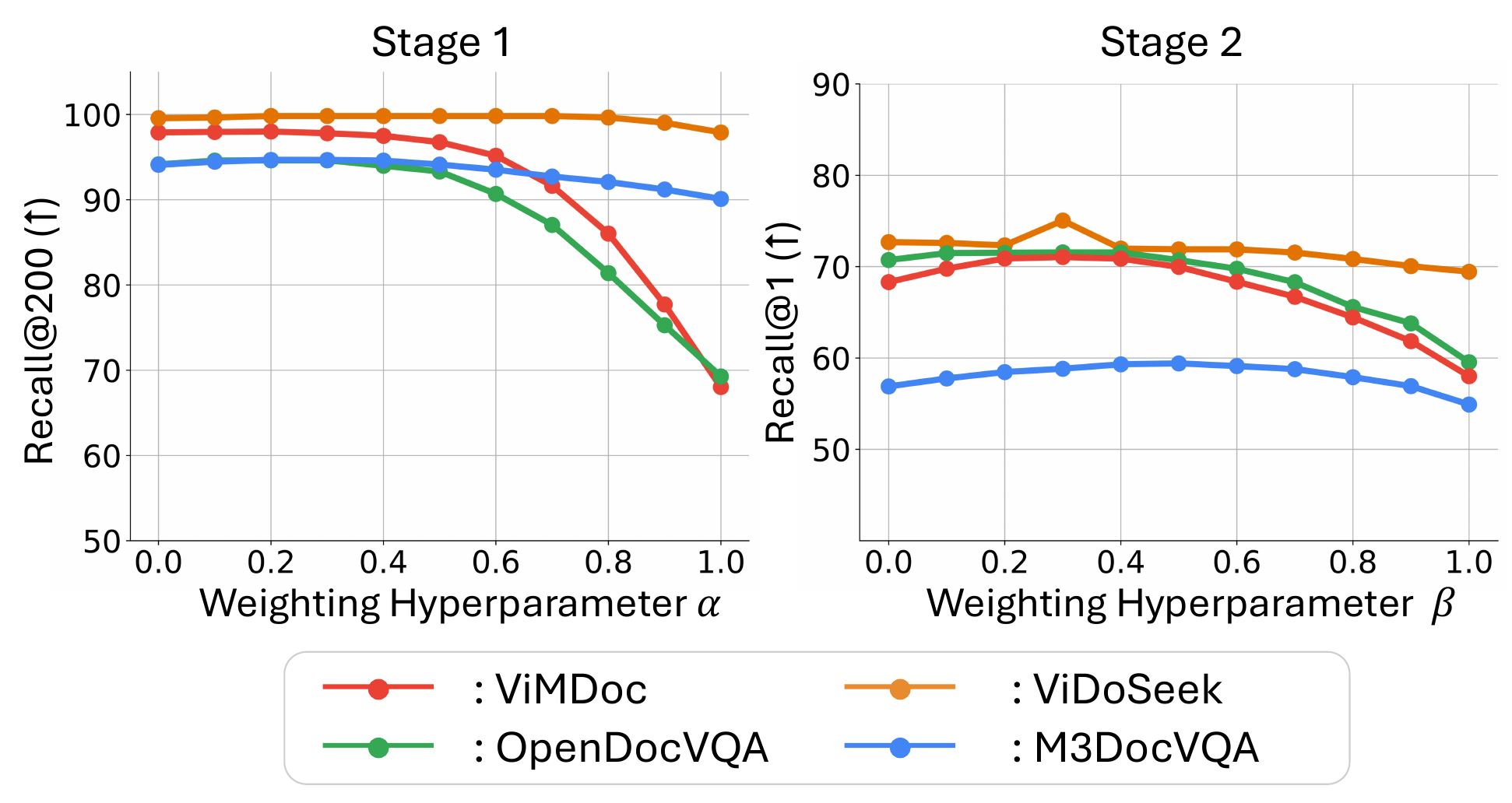}
    \vspace{-15pt}
    \caption{Hyperparameter analysis for weighting hyperparameters $\alpha$ and $\beta$.}
    \label{fig:hyper-anlysis-weight}
    \vspace{-10pt}
\end{figure}

We analyze the sensitivity of \method to hyperparameters: the reduction factor $r$ and the filtering ratios for each stage, $p_1$ and $p_2$, and the weighting hyperparameters $\alpha$ and $\beta$. 
Figure~\ref{fig:hyper-anlysis} confirms that \method remains robust across different filtering ratios in both stages, as well as across varying reduction factor values. 
Notably, \vsps are particularly effective in \benchmark, which consists of long documents.  Thus, the impact of the reduction factor is more pronounced than in other benchmarks with shorter documents (e.g., OpenDocVQA).
Figure~\ref{fig:hyper-anlysis-weight} shows that \method is also robust to
the weighting hyperparameters $\alpha$ and $\beta$ for stage 1 and 2,
respectively.

\subsection{Plug-and-Play Analysis}
\method is designed to be modular and plug-and-play, allowing the single-vector 
(Stage 1) and multi-vector (Stage 2) encoders to be replaced with other 
pretrained models. Table~\ref{tab:plug-and-play} reports results across combinations of two Stage 1 and three Stage 2 base encoders.
Among the evaluated combinations, DSE (Stage 1) and ColQwen2.5 (Stage 2) achieve the best overall performance on both \benchmark and OpenDocVQA, and are therefore adopted as the default configuration.

\begin{table}[t]

\begin{adjustbox}{max width=1.0\columnwidth}
\begin{tabular}{cc|ccc|ccc}
\toprule
\multicolumn{2}{c|}{\textbf{Base Encoder}} & \multicolumn{3}{c|}{\textbf{\benchmark} (Proposed)}  & \multicolumn{3}{c}{\textbf{OpenDocVQA}}      \\
\cmidrule(lr){1-2}
\cmidrule(lr){3-5}
\cmidrule(lr){6-8}
\textbf{Stage1}   & \textbf{Stage2}       & \textbf{R@1}   & \textbf{R@3} & \textbf{FLOPs} & \textbf{R@1} & \textbf{R@3} & \textbf{FLOPs} \\
\midrule
VisRAG & ColPali    & 60.95 & 79.19 & 1.298 & 65.27 & 80.02 & 1.287 \\
VisRAG & ColQwen2   & 62.45 & 80.07 & 0.553 & 67.79 & 81.28 & 0.615 \\
VisRAG & ColQwen2.5 & 64.06 & 82.07 & 0.553 & 67.54 & 81.40 & 0.615 \\
\midrule
DSE    & ColPali    & 63.46 & 80.68 & 1.231 & 67.50 & 81.81 & 1.214 \\
DSE    & ColQwen2   & 66.63 & 82.63 & 0.486 & 69.91 & 83.31 & 0.541 \\
DSE      & ColQwen2.5   & 71.05 & 86.41        & 0.486          & 71.56        & 84.53        & 0.541     \\    
\bottomrule
\end{tabular}
\end{adjustbox}
\caption{Plug-and-play analysis of \method with different combinations of
Stage 1 and Stage 2 base encoders on \benchmark and OpenDocVQA. R@\{1,3\} and per-query FLOPs (billions) are reported.}
\label{tab:plug-and-play}
\vspace{-5pt}
\end{table}

\section{Related Work}
\label{sec:related}

\paragraph{Visual Document Retrieval.}
Visual document retrieval matches queries against document image corpora. 
Recent methods based on LVLMs~\cite{jiang2024vlm2vec,wei2024uniir,lin2024mm}, directly encode entire document pages as images, bypassing OCR~\cite{meng2025vlm2vec,zhang2025bridging,ma2024unifying}. 
Concurrently, visual RAG leverages such visual embeddings, while multimodal approaches combine visual and textual signals using OCR~\cite{yuvisrag,suri2024visdom,wang2025vidorag,sun2025unveil} or textual summaries from LVLMs and LLMs~\cite{jain2025simpledoc,gong2025mmrag}.
In contrast, \method introduces visually-summarized pages (\vsps), an OCR-free method that enables fully-visual indexing.

\paragraph{Multi-Vector Retrieval.}
Building on the late-interaction mechanism of ColBERT~\cite{khattab2020colbert}, recent work has extended multi-vector retrieval to VDR~\cite{fayssecolpali,xu2025llama,xiao2025metaembed}. 
Although accurate, these models incur high computational overhead. 
Textual domain work addresses this via compression~\cite{santhanam2021colbertv2}, approximation~\cite{jayaram2024muvera}, and token reduction via pruning~\cite{santhanam2022plaid,acquavia2023static} or pooling~\cite{clavie2024reducing}, while VDR methods apply patch pruning~\cite{yan2025docpruner} and pooling~\cite{fayssecolpali,ma2025towards} to enhance efficiency with minimal loss in accuracy.
In contrast, \method reduces computation by filtering query tokens while maintaining performance.

\section{Conclusion}
\label{sec:conc}
We propose \method, a plug-and-play hybrid-vector framework that bridges the efficiency-accuracy trade-off in VDR.
By combining single-vector retrieval over \vsps with multi-vector reranking using selective query tokens, \method achieves near state-of-the-art accuracy with over 99\% lower computation. 
We also present \benchmark, a benchmark for visually rich, multi-document, and long-document retrieval. 
Together, they establish a scalable foundation for efficient and accurate VDR.

\section*{Limitations}
\label{sec:limit}
While \method substantially improves efficiency in visual document retrieval, several limitations remain.
First, it relies on pretrained vision-language encoders, and its performance may vary with model scale and domain adaptation quality. 
Second, the visually-summarized pages used in Stage 1 depend on document layout analysis, which can be sensitive to noisy or irregular layouts.
Third, \method focuses on retrieval efficiency and does not yet integrate retrieval augmentation generation, which we leave as future work.

\section*{Acknowledgements}
This work was supported by Samsung Electronics Co., Ltd. (IO251103-13845-01). This work was partly supported by Institute of Information \& Communications Technology Planning \& Evaluation (IITP) grant funded by the Korea government (MSIT) (No. RS-2019-II190075, Artificial Intelligence Graduate School Program (KAIST)).


\newpage
\bibliography{custom}

@inproceedings{fayssecolpali,
  title={ColPali: Efficient Document Retrieval with Vision Language Models},
  author={Faysse, Manuel and Sibille, Hugues and Wu, Tony and Omrani, Bilel and Viaud, Gautier and HUDELOT, CELINE and Colombo, Pierre},
  booktitle={ICLR},
year={2025}
}

@article{zhao2024doclayout,
  title={Doclayout-yolo: Enhancing document layout analysis through diverse synthetic data and global-to-local adaptive perception},
  author={Zhao, Zhiyuan and Kang, Hengrui and Wang, Bin and He, Conghui},
  journal={arXiv preprint arXiv:2410.12628},
  year={2024}
}

@inproceedings{khattab2020colbert,
  title={Colbert: Efficient and effective passage search via contextualized late interaction over bert},
  author={Khattab, Omar and Zaharia, Matei},
  booktitle={SIGIR},
  year={2020}
}

@inproceedings{santhanam2021colbertv2,
  title={Colbertv2: Effective and efficient retrieval via lightweight late interaction},
  author={Santhanam, Keshav and Khattab, Omar and Saad-Falcon, Jon and Potts, Christopher and Zaharia, Matei},
  booktitle={NAACL},
  year={2022}
}

@article{lee2024nv,
  title={NV-Embed: Improved Techniques for Training LLMs as Generalist Embedding Models},
  author={Lee, Chankyu and Roy, Rajarshi and Xu, Mengyao and Raiman, Jonathan and Shoeybi, Mohammad and Catanzaro, Bryan and Ping, Wei},
  journal={arXiv preprint arXiv:2405.17428},
  year={2024}
}

@inproceedings{chen2024bge,
  title={Bge m3-embedding: Multi-lingual, multi-functionality, multi-granularity text embeddings through self-knowledge distillation},
  author={Chen, Jianlv and Xiao, Shitao and Zhang, Peitian and Luo, Kun and Lian, Defu and Liu, Zheng},
  booktitle={ACL (Findings)},
  year={2024}
}

@article{tito2023hierarchical,
  title={Hierarchical multimodal transformers for multipage docvqa},
  author={Tito, Rub{\`e}n and Karatzas, Dimosthenis and Valveny, Ernest},
  journal={Pattern Recognition},
  year={2023},
  publisher={Elsevier}
}

@inproceedings{van2023document,
  title={Document understanding dataset and evaluation (dude)},
  author={Van Landeghem, Jordy and Tito, Rub{\`e}n and Borchmann, {\L}ukasz and Pietruszka, Micha{\l} and Joziak, Pawel and Powalski, Rafal and Jurkiewicz, Dawid and Coustaty, Micka{\"e}l and Anckaert, Bertrand and Valveny, Ernest and others},
  booktitle={ICCV},
  year={2023}
}

@inproceedings{tanaka2023slidevqa,
  title={Slidevqa: A dataset for document visual question answering on multiple images},
  author={Tanaka, Ryota and Nishida, Kyosuke and Nishida, Kosuke and Hasegawa, Taku and Saito, Itsumi and Saito, Kuniko},
  booktitle={AAAI},
  year={2023}
}

@article{chen2025visr,
  title={VisR-Bench: An Empirical Study on Visual Retrieval-Augmented Generation for Multilingual Long Document Understanding},
  author={Chen, Jian and Li, Ming and Kil, Jihyung and Wang, Chenguang and Yu, Tong and Rossi, Ryan and Zhou, Tianyi and Chen, Changyou and Zhang, Ruiyi},
  journal={arXiv preprint arXiv:2508.07493},
  year={2025}
}

@inproceedings{cho2024m3docrag,
  title={M3docrag: Multi-modal retrieval is what you need for multi-page multi-document understanding},
  author={Cho, Jaemin and Mahata, Debanjan and Irsoy, Ozan and He, Yujie and Bansal, Mohit},
  booktitle={ICCV Workshop},
  year={2025}
}

@inproceedings{dong2025mmdocir,
  title={Mmdocir: Benchmarking multi-modal retrieval for long documents},
  author={Dong, Kuicai and Chang, Yujing and Goh, Xin Deik and Li, Dexun and Tang, Ruiming and Liu, Yong},
  booktitle={EMNLP},
  year={2025}
}

@inproceedings{ma2024mmlongbench,
  title={Mmlongbench-doc: Benchmarking long-context document understanding with visualizations},
  author={Ma, Yubo and Zang, Yuhang and Chen, Liangyu and Chen, Meiqi and Jiao, Yizhu and Li, Xinze and Lu, Xinyuan and Liu, Ziyu and Ma, Yan and Dong, Xiaoyi and others},
  booktitle={NeurIPS},
  year={2024}
}

@inproceedings{tanaka2025vdocrag,
  title={Vdocrag: Retrieval-augmented generation over visually-rich documents},
  author={Tanaka, Ryota and Iki, Taichi and Hasegawa, Taku and Nishida, Kyosuke and Saito, Kuniko and Suzuki, Jun},
  booktitle={CVPR},
  year={2025}
}

@inproceedings{jain2025simpledoc,
  title={SimpleDoc: Multi-Modal Document Understanding with Dual-Cue Page Retrieval and Iterative Refinement},
  author={Jain, Chelsi and Wu, Yiran and Zeng, Yifan and Liu, Jiale and Shao, Zhenwen and Wu, Qingyun and Wang, Huazheng and others},
  booktitle={EMNLP},
  year={2025}
}

@inproceedings{zhang2025bridging,
  title={Bridging Modalities: Improving Universal Multimodal Retrieval by Multimodal Large Language Models},
  author={Zhang, Xin and Zhang, Yanzhao and Xie, Wen and Li, Mingxin and Dai, Ziqi and Long, Dingkun and Xie, Pengjun and Zhang, Meishan and Li, Wenjie and Zhang, Min},
  booktitle={CVPR},
  year={2025}
}

@article{gong2025mmrag,
  title={MMRAG-DocQA: A Multi-Modal Retrieval-Augmented Generation Method for Document Question-Answering with Hierarchical Index and Multi-Granularity Retrieval},
  author={Gong, Ziyu and Huang, Yihua and Mai, Chengcheng},
  journal={arXiv preprint arXiv:2508.00579},
  year={2025}
}

@inproceedings{wang2025vidorag,
  title={Vidorag: Visual document retrieval-augmented generation via dynamic iterative reasoning agents},
  author={Wang, Qiuchen and Ding, Ruixue and Chen, Zehui and Wu, Weiqi and Wang, Shihang and Xie, Pengjun and Zhao, Feng},
  booktitle={EMNLP},
  year={2025}
}

@article{xu2025llama,
  title={Llama Nemoretriever Colembed: Top-Performing Text-Image Retrieval Model},
  author={Xu, Mengyao and Moreira, Gabriel and Ak, Ronay and Osmulski, Radek and Babakhin, Yauhen and Yu, Zhiding and Schifferer, Benedikt and Oldridge, Even},
  journal={arXiv preprint arXiv:2507.05513},
  year={2025}
}

@inproceedings{yuvisrag,
  title={VisRAG: Vision-based Retrieval-augmented Generation on Multi-modality Documents},
  author={Yu, Shi and Tang, Chaoyue and Xu, Bokai and Cui, Junbo and Ran, Junhao and Yan, Yukun and Liu, Zhenghao and Wang, Shuo and Han, Xu and Liu, Zhiyuan and others},
  booktitle={ICLR},
year={2024}
}

@inproceedings{wasserman2025real,
  title={REAL-MM-RAG: A Real-World Multi-Modal Retrieval Benchmark},
  author={Wasserman, Navve and Pony, Roi and Naparstek, Oshri and Goldfarb, Adi Raz and Schwartz, Eli and Barzelay, Udi and Karlinsky, Leonid},
  booktitle={ACL},
  year={2025}
}

@inproceedings{deng2024longdocurl,
  title={Longdocurl: a comprehensive multimodal long document benchmark integrating understanding, reasoning, and locating},
  author={Deng, Chao and Yuan, Jiale and Bu, Pi and Wang, Peijie and Li, Zhong-Zhi and Xu, Jian and Li, Xiao-Hui and Gao, Yuan and Song, Jun and Zheng, Bo and others},
  booktitle={ACL},
  year={2025}
}

@inproceedings{ma2024unifying,
  title={Unifying multimodal retrieval via document screenshot embedding},
  author={Ma, Xueguang and Lin, Sheng-Chieh and Li, Minghan and Chen, Wenhu and Lin, Jimmy},
  booktitle={EMNLP},
  year={2024}
}

@inproceedings{suri2024visdom,
  title={Visdom: Multi-document qa with visually rich elements using multimodal retrieval-augmented generation},
  author={Suri, Manan and Mathur, Puneet and Dernoncourt, Franck and Goswami, Kanika and Rossi, Ryan A and Manocha, Dinesh},
  booktitle={NAACL},
  year={2025}
}

@inproceedings{sun2025unveil,
  title={Unveil: Unified Visual-Textual Integration and Distillation for Multi-modal Document Retrieval},
  author={Sun, Hao and Hou, Yingyan and Guo, Jiayan and Wang, Bo and Yang, Chunyu and Ni, Jinsong and Zhang, Yan},
  booktitle={ACL},
  year={2025}
}

@inproceedings{ma2025towards,
  title={Towards Storage-Efficient Visual Document Retrieval: An Empirical Study on Reducing Patch-Level Embeddings},
  author={Ma, Yubo and Li, Jinsong and Zang, Yuhang and Wu, Xiaobao and Dong, Xiaoyi and Zhang, Pan and Cao, Yuhang and Duan, Haodong and Wang, Jiaqi and Cao, Yixin and others},
  booktitle={ACL (Findings)},
  year={2025}
}

@inproceedings{qiao2025reproducibility,
  title={Reproducibility, Replicability, and Insights into Visual Document Retrieval with Late Interaction},
  author={Qiao, Jingfen and Ju, Jia-Huei and Ma, Xinyu and Kanoulas, Evangelos and Yates, Andrew},
  booktitle={SIGIR},
  year={2025}
}

@article{clavie2024reducing,
  title={Reducing the footprint of multi-vector retrieval with minimal performance impact via token pooling},
  author={Clavi{\'e}, Benjamin and Chaffin, Antoine and Adams, Griffin},
  journal={arXiv preprint arXiv:2409.14683},
  year={2024}
}

@inproceedings{santhanam2022plaid,
  title={PLAID: an efficient engine for late interaction retrieval},
  author={Santhanam, Keshav and Khattab, Omar and Potts, Christopher and Zaharia, Matei},
  booktitle={CIKM},
  year={2022}
}

@article{yan2025docpruner,
  title={DocPruner: A Storage-Efficient Framework for Multi-Vector Visual Document Retrieval via Adaptive Patch-Level Embedding Pruning},
  author={Yan, Yibo and Xu, Guangwei and Zou, Xin and Liu, Shuliang and Kwok, James and Hu, Xuming},
  journal={arXiv preprint arXiv:2509.23883},
  year={2025}
}

@inproceedings{smith2007overview,
  title={An overview of the Tesseract OCR engine},
  author={Smith, Ray},
  booktitle={ICDAR},
  year={2007},
}

@article{xiao2025metaembed,
  title={MetaEmbed: Scaling Multimodal Retrieval at Test-Time with Flexible Late Interaction},
  author={Xiao, Zilin and Ma, Qi and Gu, Mengting and Chen, Chun-cheng Jason and Chen, Xintao and Ordonez, Vicente and Mohan, Vijai},
  journal={arXiv preprint arXiv:2509.18095},
  year={2025}
}

@inproceedings{acquavia2023static,
  title={Static pruning for multi-representation dense retrieval},
  author={Acquavia, Antonio and Macdonald, Craig and Tonellotto, Nicola},
  booktitle={DocEng},
  year={2023}
}

@inproceedings{jayaram2024muvera,
  title={MUVERA: Multi-Vector Retrieval via Fixed Dimensional Encoding},
  author={Jayaram, Rajesh and Dhulipala, Laxman and Hadian, Majid and Lee, Jason D and Mirrokni, Vahab},
  booktitle={NeurIPS},
  year={2024}
}

@inproceedings{jiang2024vlm2vec,
  title={Vlm2vec: Training vision-language models for massive multimodal embedding tasks},
  author={Jiang, Ziyan and Meng, Rui and Yang, Xinyi and Yavuz, Semih and Zhou, Yingbo and Chen, Wenhu},
  booktitle={ICLR},
  year={2025}
}

@article{meng2025vlm2vec,
  title={Vlm2vec-v2: Advancing multimodal embedding for videos, images, and visual documents},
  author={Meng, Rui and Jiang, Ziyan and Liu, Ye and Su, Mingyi and Yang, Xinyi and Fu, Yuepeng and Qin, Can and Chen, Zeyuan and Xu, Ran and Xiong, Caiming and others},
  journal={arXiv preprint arXiv:2507.04590},
  year={2025}
}

@inproceedings{wei2024uniir,
  title={Uniir: Training and benchmarking universal multimodal information retrievers},
  author={Wei, Cong and Chen, Yang and Chen, Haonan and Hu, Hexiang and Zhang, Ge and Fu, Jie and Ritter, Alan and Chen, Wenhu},
  booktitle={ECCV},
  year={2024},
}

@inproceedings{lin2024mm,
  title={Mm-embed: Universal multimodal retrieval with multimodal llms},
  author={Lin, Sheng-Chieh and Lee, Chankyu and Shoeybi, Mohammad and Lin, Jimmy and Catanzaro, Bryan and Ping, Wei},
booktitle={ICLR},
  year={2025}
}

\newpage
\appendix
\label{sec:appendix}
\section*{Appendix}
\label{sec:app}

\section{\method}

\subsection{Detailed \vs Construction}
\label{app:vs-page-detail}
For Document Layout Detection ($\mathsf{DLA}$), we utilized DocLayout-YOLO~\cite{zhao2024doclayout}\footnote{\url{juliozhao/DocLayout-YOLO-DocStructBenct/doclayout_yolo_docstructbench_imgsz1024.pt}}, which classifies layouts into 8 categories: title, plain text, abandon, figure, figure caption, table, table caption, isolated formula, and formula caption. To construct the \vs, only the title layouts were extracted by cropping each detected title region according to its bounding box. Following extraction, the cropped title images were assembled into a single page via vertical stacking, a process defined as $\mathsf{Assemble(\cdot)}$. No resizing or scaling operations were applied to preserve the original visual context. 

\subsection{\vs Examples}
\label{app:vs-page-example}
\begin{figure*}[!ht]
    \centering
    \includegraphics[width=0.95\linewidth]{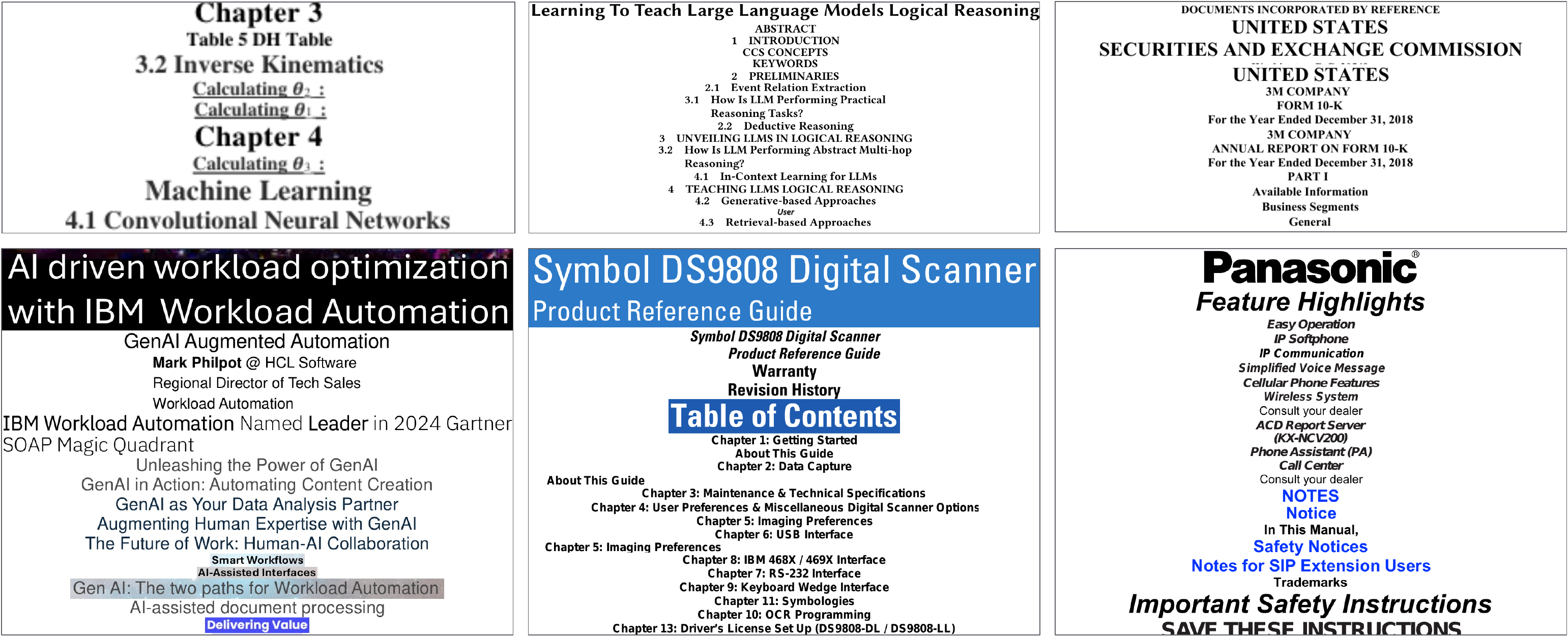}
    \caption{\vs examples from \benchmark.}
    \label{fig:vs-vimdoc}
    \vspace{5mm}
\end{figure*}

\begin{figure*}[!ht]
    \centering
    \includegraphics[width=0.95\linewidth]{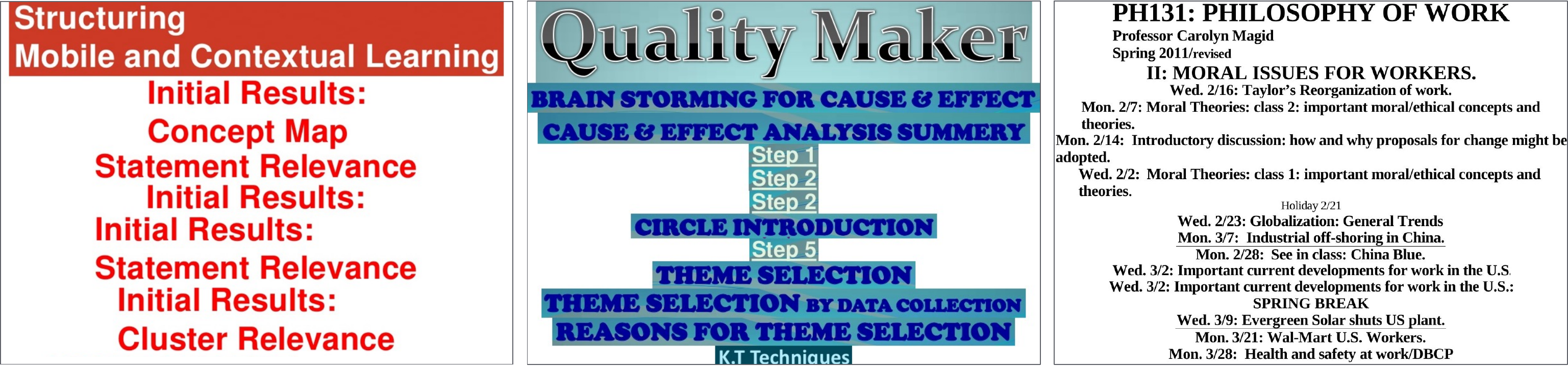}
    \caption{\vs examples from OpenDocVQA.}
    \label{fig:vs-opendocvqa}
    \vspace{5mm}
\end{figure*}

\begin{figure*}[!ht]
    \centering
    \includegraphics[width=0.95\linewidth]{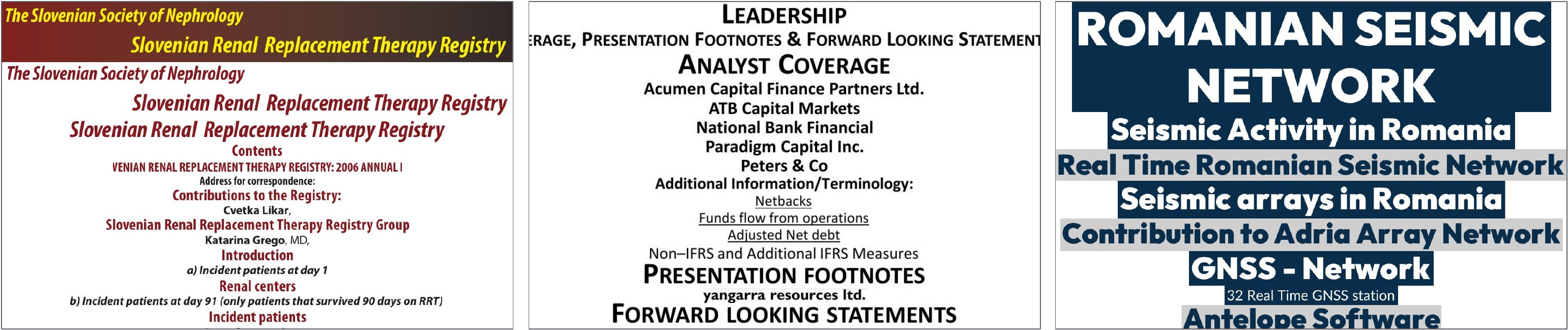}
    \caption{\vs examples from ViDoSeek.}
    \label{fig:vs-vidoseek}
    \vspace{5mm}
\end{figure*}

\begin{figure*}[!ht]
    \centering
    \includegraphics[width=0.95\linewidth]{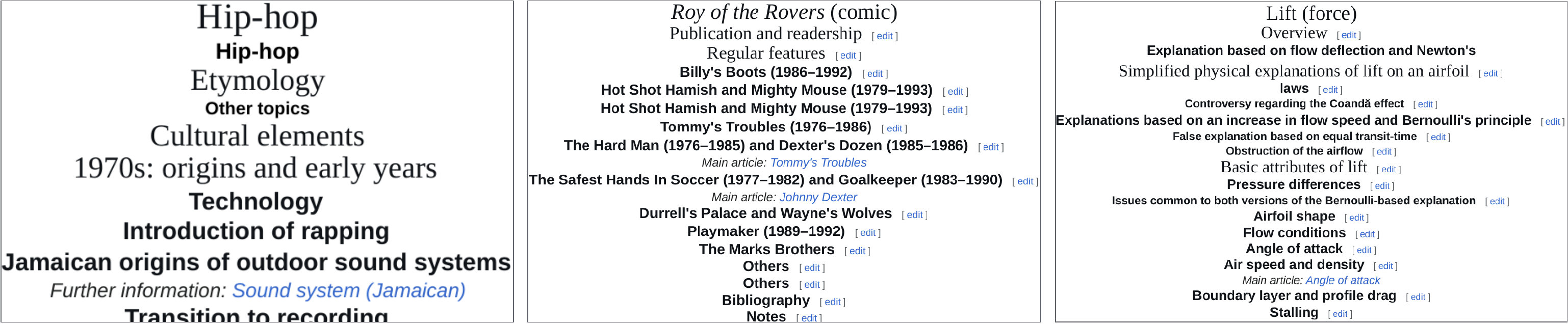}
    \caption{\vs examples from M3DocVQA.}
    \label{fig:vs-m3docvqa}
     \vspace{15mm}
\end{figure*}

In this section, we further provide more examples of the constructed \vs. Figure~\ref{fig:vs-vimdoc}, \ref{fig:vs-opendocvqa}, \ref{fig:vs-vidoseek}, and \ref{fig:vs-m3docvqa} show example output across the benchmarks.

\subsection{Detailed Query \textit{Key Token} Filtering}
\label{app:method-key-token}
Part-of-Speech tagging is used to identify the subset of \textit{key tokens}. \texttt{nltk}\footnote{We use \texttt{nltk} 3.9.2 version.} is used for Part-of-Speech(POS) tagging. Using the tokenized query from the model, tokens with these four tags are filtered as \textit{key tokens}: NN, NNS, NNP, and NNPS.  

\subsection{\textit{Key Token} Distribution}
Table~\ref{tab:token-distribution} details the token distribution per benchmark. \textit{Key tokens} comprise only 30-37\% of all query tokens, with the remainder being non-\textit{key tokens}.
\label{app:method-key-distribution}
\begin{table}[h]

\begin{adjustbox}{max width=1.0\columnwidth}
\begin{tabular}{l|cl|cl}
\toprule
\textbf{Benchmarks} & \textbf{\#Avg. \textit{key}} & \textbf{( \% )} & \textbf{\#Avg. non-\textit{key}} & \textbf{( \% )} \\
\midrule
\textbf{\benchmark}       & 6.8  $\pm$ 4.1 & ( 30.32\% ) & 15.5 $\pm$ 7.9 & ( 69.68\% ) \\
\textbf{OpenDocVQA} & 6.6  $\pm$ 3.4 & ( 31.51\% ) & 14.2 $\pm$ 7.6 & ( 68.49\% )\\
\textbf{ViDoSeek}   & 11.2 $\pm$ 3.9 & ( 37.48\% ) & 18.7 $\pm$ 5.8 & ( 62.52\% ) \\
\textbf{M3DocVQA}   & 9.1  $\pm$ 5.2 & ( 34.63\% ) & 17.2 $\pm$ 8.6 & ( 65.37\% )\\
\midrule
\rowcolor[HTML]{EFEFEF}
\textbf{Average}  & 7.4  $\pm$ 4.5 &( 31.83\% )  & 15.9 $\pm$ 7.9 & ( 68.17\% )  \\ 
\bottomrule
\end{tabular}
\end{adjustbox}
\caption{Average number and percentage of \textit{key tokens} and non-\textit{key tokens} across benchmarks.}
\label{tab:token-distribution}
\end{table}

\section{\benchmark Benchmark}
\label{app:benchmark}
\subsection{Detailed Data Collection}
\label{app:bench-collection}
In this section, we provide details of the data collection process. 

\begin{itemize}[leftmargin=*,itemsep=0pt]
    \item REAL-MM-RAG~\cite{wasserman2025real}: All four splits (FinReport, FinSlides, TechSlides, and TechReport) were included.
    \item VisR-Bench~\cite{chen2025visr}: Only English splits were retained, excluding the Multilingual split, since baseline models were pre-trained mainly on English datasets. Within the English split, all three query types (figure, table, and text) were included.
    \item MMDocIR~\cite{dong2025mmdocir}: Only the evaluation set was used.  
    \item LongDocURL~\cite{deng2024longdocurl}: All documents included in the paper were used.
    \item MMLongBench-Doc~\cite{ma2024mmlongbench}: Similar to LongDocURL, all documents were included.
\end{itemize}

\subsection{Detailed Query Processing}
\label{app:bench-query}

\begin{figure*}[!ht]
    \centering
    \includegraphics[width=0.95\linewidth]{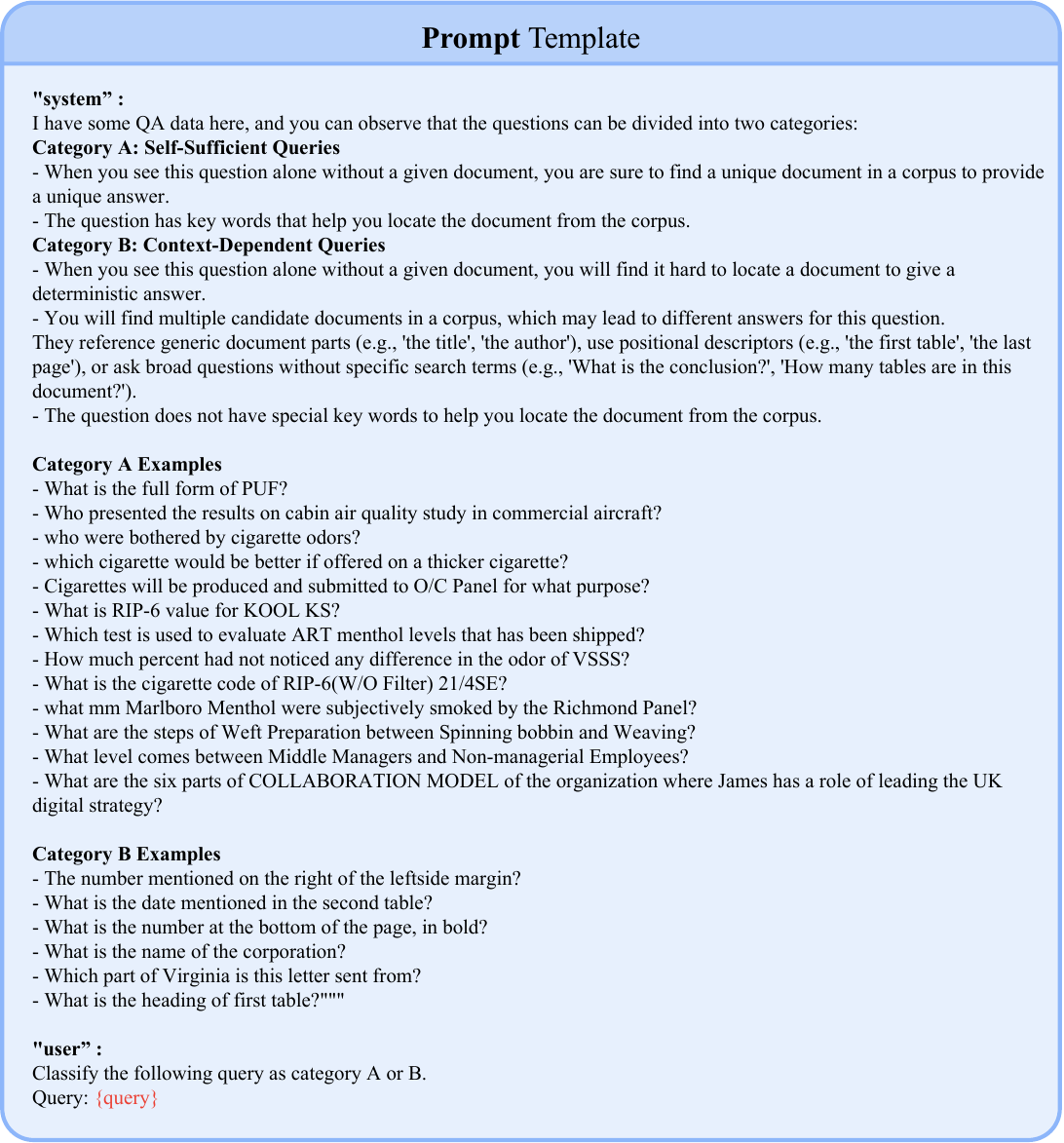}
    \caption{Prompt template for Query Filtering. Modified from Quality Reviewer prompt in \cite{wang2025vidorag}.}
    \label{fig:prompt-query-filter}
    \vspace{10mm}
\end{figure*}

Table~\ref{tab:bench-summary} shows that the two-stage filtering removes 45.8\%
of queries, retaining only \textit{self-contained} ones suitable for multi-document retrieval setting. \textbf{Heuristic rule-based filtering}
removes `Unanswerable' queries from LongDocURL~\cite{deng2024longdocurl}
and MMLongBench-Doc~\cite{ma2024mmlongbench} that are not associated with any page, and  queries that explicitly reference `Table $N$' or `Figure $N$',
which are inherently \textit{context-dependent}. However, some
\textit{context-dependent} queries lack explicit keywords and are thus
difficult to catch with heuristic rules alone. We therefore apply \textbf{LLM-based
filtering} to remove such cases. Figure~\ref{fig:prompt-query-filter} shows the prompt used for LLM-based filtering following \cite{wang2025vidorag}. We employ \texttt{gpt-5-mini} for LLM-based filtering.


\begin{table}[!ht]

\begin{adjustbox}{max width=1.0\columnwidth}
\begin{tabular}{l|cc|c}
\toprule
& \textbf{\#Avg. Page} & \textbf{\#Doc} & \textbf{\#Query} \small{(\%Filtered)} \\
\midrule
\textbf{\benchmark (Proposed)} & \textbf{55.4} & \textbf{1379} &  \textbf{10904} \small{(-45.8\%)} \\ \midrule %
REAL-MM-RAG & 52.8 &  162&  3939 \small{(-13.5\%)} \\
VisR-Bench & 18.5 & 373 & 5142 \small(-50.9\%)  \\
MMDocIR & 65.1 & 313 &  651 \small{(-60.7\%)}  \\
LongDocURL & 85.6 & 396 &  890 \small{(-61.7\%)}  \\
MMLongBench-Doc & 47.5 & 135 &  282 \small{(-73.9\%)}  \\ \bottomrule

\end{tabular}
\end{adjustbox}
\caption{Dataset splits included in \benchmark.}
\label{tab:bench-summary}
\end{table}

\section{Experimental Details}
\label{app:exp-detail}

\subsection{Models}
\label{app:exp-detail-model}

Table~\ref{tab:model-checkpoint} shows embedding dimension, base model, batch size and model checkpoints from Hugging Face used in experiments.

\begin{table*}[!ht]
\centering

\setlength{\extrarowheight}{2pt}
\scalebox{0.7}{
\begin{tabular}{cl|c|c|c|c|l}
\toprule
& \textbf{Model} & \textbf{Model Size} & \textbf{Dimension} & \textbf{Base Model} & \textbf{Batch Size} & \textbf{Checkpoint} \\ 
\midrule
\multirow{5}{*}{\rotatebox[origin=c]{90}{Single-Vector}}
& BGE-M3 (dense) \textcolor{gray}{\textsuperscript{\faFileTextO}} & 560M & 1,024 & XLMRoberta & 16 & \url{BAAI/bge-m3} \\
& NV-Embed-V2 \textcolor{gray}{\textsuperscript{\faFileTextO}} & 7.85B & 4,096 & Mistral-7B & 16 & \url{nvidia/NV-Embed-v2} \\
& VisRAG \textcolor{gray}{\textsuperscript{\faPictureO}} & 3.43B & 2,304 & MiniCPM-V-2 & 1 & \url{openbmb/VisRAG-Ret} \\
& GME \textcolor{gray}{\textsuperscript{\faPictureO}} & 2.21B & 1,536 & Qwen2-VL-2B & 1 & \url{Alibaba-NLP/gme-Qwen2-VL-2B-Instruct} \\

& DSE \textcolor{gray}{\textsuperscript{\faPictureO}} & 2.21B & 1,536 & Qwen2-VL-2B & 1 &\url{MrLight/dse-qwen2-2b-mrl-v1} \\
\midrule
\multirow{4}{*}{\rotatebox[origin=c]{90}{Multi-Vector}} & BGE-M3 (multi) \textcolor{gray}{\textsuperscript{\faFileTextO}} & 560M & $n_{token} \times 1,024$ & XLMRoberta & 16 & \url{BAAI/bge-m3} \\
& ColPali \textcolor{gray}{\textsuperscript{\faPictureO}} & 2.92B & $n_{patch} \times 128$ & PaliGemma-3B & 1  &  \url{vidore/colpali-v1.3}  \\
& ColQwen2 \textcolor{gray}{\textsuperscript{\faPictureO}} & 2.21B & $n_{patch} \times 128$ & Qwen2-VL-2B & 1 & \url{vidore/colqwen2-v1.0} \\
& ColQwen2.5 \textcolor{gray}{\textsuperscript{\faPictureO}} & 3.75B & $n_{patch} \times 128$ & Qwen2.5-VL-3B & 1 & \url{vidore/colqwen2.5-v0.1} \\ 
\bottomrule
\end{tabular}
}
\caption{Model size, output dimension, base model, batch size and checkpoint from Hugging Face model used in experiments. \textcolor{gray}{\faFileTextO} indicates a textual embedding model and \textcolor{gray}{\faPictureO} indicates a visual embedding model. }

\label{tab:model-checkpoint}
\end{table*}

\subsection{Benchmark Statistics}
\label{app:exp-detail-benchmark}
Table~\ref{tab:bench-statistics-all} provides detailed statistics for the four benchmarks used in our experiment. For OpenDocVQA, the SlideVQA and DUDE splits were merged, which consist of multi-page documents.
This follows the \textit{all-pool} setting in the original paper, but excludes the InfoVQA and ChartQA splits, which consist of single-page documents.

\begin{table}[h]

\centering
\begin{adjustbox}{max width=0.9\columnwidth}
\begin{tabular}{l|cc}
\toprule
\textbf{Benchmarks} & \textbf{\# Pages (Images)} & \textbf{\# Query} \\
\midrule
\textbf{\benchmark} (Ours)      & 76,347            & 10,904    \\
OpenDocVQA & 80,335            & 1,256     \\
ViDoSeek   & 5,385             & 1,142     \\
M3DocVQA   & 41,071            & 2,441    \\
\bottomrule
\end{tabular}
\end{adjustbox}
\caption{Statistics of the used benchmarks.}
\label{tab:bench-statistics-all}
\end{table}

\subsection{Evaluation Metric}
\label{app:exp-detail-metric}

\paragraph{Document-level Retrieval} 
Given a document $D_k = ( P_{k,1}, P_{k,2}, \ldots, P_{k,|D_k|})$,
document-level retrieval accuracy $S_\mathrm{}(q, D_k)$ is defined using the maximum value of page-level retrieval accuracy: 
\begin{equation}
S_\mathrm{}(q, D_k) = max_{P \in D_{k}} S(q,P)
\end{equation}
where $S$ can be either $S_\mathrm{SV}$ or $S_\mathrm{MV}$.

\paragraph{Recall@K}
Recall@K measures the proportion of ground truth pages that appear in the top-K retrieved results.
Given a query, let $P$ a set of all ground truth pages and $P_K$  a set of ground truth pages in the top-$K$ retrieved pages.
Recall@K is defined as:
\begin{equation}
\text{Recall@K} = \frac{|P_K|}{|P|}
\end{equation}
where $|P_K|$ is the number of ground truth pages retrieved in the top-K results, and $|P|$ is the total number of ground truth pages for the query.

\paragraph{FLOPs}
FLOPs were used as a retrieval efficiency evaluation metric along with latency in Table~\ref{tab:latency}. It was observed that token length variations in queries and documents significantly affect scoring FLOPs during batch processing due to padding. For fair comparison, a batch size of 1 is assumed for all FLOPs calculations. All embeddings were processed and compared using float16 precision.
Detailed analysis is provided in Table~\ref{tab:flops-calculation}.

\begin{table}[!ht]
\centering

\setlength{\extrarowheight}{3pt}
\begin{adjustbox}{max width=\columnwidth}
\begin{tabular}{l|l}
\toprule
\textbf{Method} & \textbf{FLOPs per Query} \\
\midrule
\textbf{Single-Vector Retrieval} & $O(d\,|\mathcal{P}|)$ \\
\textbf{Multi-Vector Retrieval} & $O(d\,n_q\sum_{P\in\mathcal{P}} n_P)$ \\
\midrule
\rowcolor[HTML]{F7F7F7}
\textbf{\method\ (Ours)} &  \\
\midrule
\quad Stage~1: Candidate Retrieval & 
$O(d\,|\mathcal{VS}|)$, where $|\mathcal{VS}|\!\approx\!|\mathcal{P}|/r$ \\
\quad Stage~1: Refinement & 
$O(d\,|\mathcal{C}|)$, where $|\mathcal{C}|\!\approx\!p_1|\mathcal{P}|$ \\
\midrule
\quad Stage~2: Reranking & 
$O(d\,|q_\text{key}|\sum_{P\in\mathcal{C}_K}n_P)$, where $|q_\text{key}|\!\approx\!0.3\,n_q$, $|\mathcal{C}_K|\!=\!K$ \\
\quad Stage~2: Refinement & 
$O(d\,n_q\sum_{P\in\mathcal{C}^\star}n_P)$, where $|\mathcal{C}^\star|\!=\!p_2K$ \\
\bottomrule
\end{tabular}
\end{adjustbox}
\caption{Computation cost analysis (FLOPs per query).}
\label{tab:flops-calculation}
\end{table}

\subsection{Hyperparameter for Efficiency Variants}
\label{app:exp-detail-eff}

We detail the specific hyperparameter ranges used to scale the existing efficiency variants of multi-vector models.

\begin{itemize}[leftmargin=*]
    \item \textbf{\method:} $C_k$, which defines the number of refine candidates from Stage~1, is used for scaling and tested using the set $C_k = \{100, 200, 400\}$.
    \item \textbf{Document Patch Pooling:} Hierarchical Pooler from ColPali~\cite{fayssecolpali}\footnote{\url{https://pypi.org/project/colpali-engine/}} is used for patch pooling. Pool factors are varied across the range from $2^2$ (4 patches) to $12^2$ (144 patches).
    \item \textbf{Document Patch Pruning:} Following ~\cite{ma2025towards,yan2025docpruner}, random pruning is employed as a baseline, which has shown superior performance descpite its simplicity.
    Pruning ratios are set from $0.1$ to $0.9$ (using $0.1$ increments). Additional fine-grained ratios of $\{0.075, 0.095, 0.925\}$ are included to allow for comparison at similar Floating Point Operations (FLOPs) scales.
    \setcounter{enumi}{2}
    \item \textbf{Query Token Pooling:} The same pooler used for document patch pooling is utilzed. Considering the variance in query token length, pooling are performed on the fixed-length query special tokens\footnote{Also referred to as query augmentation tokens.}, which has proven effective in~\cite{qiao2025reproducibility,fayssecolpali}. Pool factors  of $\{2, 3, 5, 10\}$ are selected.
    \item \textbf{Query Token Pruning:} The same random pruning method is applied to the query augmentation tokens. Pruning ratios range from $0.1$ to $0.9$ (using $0.1$ increments).
\end{itemize}

\subsection{Extended Experimental Results}
\label{app:exp-extended}
\begin{table*}[h]
\centering

\begin{adjustbox}{max width=1.8\columnwidth}
\begin{tabular}{l|c|l|c|ccc|ccc}
\toprule

& \multirow{2}{*}{\textbf{\#Cross-Page}}  &        &                & \multicolumn{3}{c}{\textbf{Cross-page Query}} & \multicolumn{3}{|c}{\textbf{Single-page Query}} \\
\cmidrule(lr){5-7} \cmidrule(lr){8-10}
& \textbf{Query (\%)} & \textbf{Methods} & \textbf{FLOPs} & \textbf{R@1}  & \textbf{R@3}  & \textbf{R@5}  & \textbf{R@1}   & \textbf{R@3}  & \textbf{R@5}  \\
\midrule
\benchmark      & 677            & DSE              & 0.235          & 23.21         & 44.76         & 55.10         & 60.33          & 79.22         & 84.74         \\
&           (6.2\%)     & \textbf{\method(only Stage 1)}             & 0.134          & 22.65         & 44.07         & 54.86         & 59.97          & 78.71         & 84.28         \\
\cmidrule(lr){3-10}
&                & ColQwen2.5       & 407.320        & 25.92         & 50.25         & 60.13         & 74.10          & 88.77         & 92.30         \\
&                & \textbf{\method}             & 0.486          & 25.22         & 48.16         & 57.83         & 73.58          & 88.38         & 91.90         \\
\midrule
OpenDocVQA & 259            & DSE              & 0.247          & 36.97         & 64.96         & 73.58         & 65.20          & 78.64         & 82.95         \\
&       (10.6\%)         & \textbf{\method(only Stage 1)}             & 0.147          & 37.93         & 65.35         & 74.68         & 65.60          & 78.84         & 82.95         \\
\cmidrule(lr){3-10}
&                & ColQwen2.5       & 482.049        & 42.18         & 72.94         & 79.18         & 80.54          & 89.87         & 91.78         \\
&                & \textbf{\method}             & 0.541          & 43.15         & 72.68         & 78.22         & 77.73          & 86.76         & 88.67        \\

 \bottomrule
\end{tabular}
\end{adjustbox}
\caption{Experimental results for Cross-page Query (requiring retrieval of multiple pages) and Single-page Query. For the Cross-page Query, the average number of pages to be retrieved is 2.4, 2.0 for \benchmark, OpenDocVQA, respectively. Recall@\{1, 3, 5\} is reported for both query types.}
\label{tab:cross-query}
\end{table*}

In this section, we provide further experimental results. Table~\ref{tab:cross-query} shows detailed results of cross-page query, which requires more than one page to be retrieved, and single-page query results. Following ~\cite{tanaka2025vdocrag}, Table~\ref{tab:main-split-vimdoc} and Table~\ref{tab:main-split-opendoc} show experimental results where each retrieval is performed under each data split.

\begin{table*}[!ht]
\centering

\setlength\tabcolsep{1.7pt} 
\setlength{\extrarowheight}{2pt}
\scalebox{0.67}{
\begin{tabular}{cl|ccc|ccc|ccc|ccc|ccc}
\toprule
& & \multicolumn{15}{c}{\textbf{\large{\benchmark}} (Proposed)}  \\
\cmidrule(lr){3-17} 
& \textbf{Data Split} & \multicolumn{3}{c|}{\textbf{REAL-MM-RAG}} & \multicolumn{3}{c|}{\textbf{VisR-Bench}} & \multicolumn{3}{c|}{\textbf{MMDocIR}} & \multicolumn{3}{c|}{\textbf{LongDocURL}} & \multicolumn{3}{c}{\textbf{MMLongBench-Doc}} \\
\midrule
& \textbf{Model} & \textbf{R@1} & \textbf{R@3} & \textbf{FLOPs} & \textbf{R@1} & \textbf{R@3} & \textbf{FLOPs} & \textbf{R@1} & \textbf{R@3} & \textbf{FLOPs} & \textbf{R@1} & \textbf{R@3} & \textbf{FLOPs} & \textbf{R@1} & \textbf{R@3} & \textbf{FLOPs}  \\
\midrule

\multirow{5}{*}{\rotatebox[origin=c]{90}{Single-Vector}} & 
 BGE-M3 (dense) \textcolor{gray}{\textsuperscript{\faFileTextO}}& 38.69 & 56.13 & 0.018 & 58.75 & 75.32 & 0.014 & 31.49 & 46.06 & 0.042 & 30.89 & 46.32 & 0.069 & 39.95 & 60.34 & 0.013 \\
& NV-Embed-V2 \textcolor{gray}{\textsuperscript{\faFileTextO}}& 49.48 & 68.52 & 0.070 & 59.98 & 79.52 & 0.057 & 35.28 & 53.66 & 0.167 & 35.58 & 54.66 & 0.278 & 45.79 & 69.99 & 0.053 \\

& VisRAG \textcolor{gray}{\textsuperscript{\faPictureO}} & 36.56 & 53.87 & 0.040 & 60.79 & 81.10 & 0.032 & 40.13 & 62.62 & 0.094 & 41.87 & 61.98 & 0.156 & 41.34 & 61.58 & 0.030 \\

& GME \textcolor{gray}{\textsuperscript{\faPictureO}} & 53.31 & 73.72 & 0.026 & 68.86 & 86.31 & 0.021 & 50.44 & 69.74 & 0.063 & 44.78 & 67.19 & 0.104 & 51.28 & 69.62 & 0.020 \\

& DSE \textcolor{gray}{\textsuperscript{\faPictureO}} & 53.74 & 73.85 & 0.026 & 69.88 & 87.07 & 0.021 & 52.71 & 71.02 & 0.063 & 45.08 & 66.96 & 0.104 & 45.07 & 66.40 & 0.020 \\ 
\midrule
\rowcolor[HTML]{EFEFEF} 

& \textbf{\method (only Stage 1)} & 53.82 & 73.80 & 0.015 & 68.92 & 85.53 & 0.013 & 52.10 & 70.10 & 0.036 & 44.41 & 66.26 & 0.059 & 45.42 & 65.87 & 0.011 \\
& (vs. DSE) & \sblue{100.14\%} & \sblue{99.93\%} & \sred{-42.61\%} & \sblue{98.64\%} & \sblue{98.24\%} & \sred{-40.40\%} & \sblue{98.83\%} & \sblue{98.70\%} & \sred{-42.97\% }& \sblue{98.51\%} & \sblue{98.95\%} & \sred{-43.13\%} & \sblue{100.79\%} & {99.20\%} & \sred{-42.92\%}  \\
\midrule

\multirow{4}{*}{\rotatebox[origin=c]{90}{Multi-Vector}} &
BGE-M3 (multi) \textcolor{gray}{\textsuperscript{\faFileTextO}} & 48.67 & 66.03 & 157.599 & 68.26 & 84.23 & 101.84 & 32.89 & 49.24 & 491.229 & 33.34 & 50.95 & 837.209 & 43.11 & 65.10 & 164.000 \\

& Colpali \textcolor{gray}{\textsuperscript{\faPictureO}} & 59.25 & 77.61 & 75.914 & 75.83 & 90.22 & 55.602 & 50.29 & 67.84 & 189.206 & 47.69 & 69.79 & 402.024 & 43.26 & 62.94 & 68.038 \\
& ColQwen2 \textcolor{gray}{\textsuperscript{\faPictureO}} & 60.88 & 77.43 & 53.484 & 80.47 & 92.82 & 39.549 & 54.80 & 71.72 & 109.115 & 51.90 & 69.67 & 227.017 & 48.53 & 69.17 & 48.420 \\
& ColQwen2.5 \textcolor{gray}{\textsuperscript{\faPictureO}} & 68.85 & 85.61 & 53.484 & 81.95 & 93.27 & 39.549 & 59.41 & 77.06 & 109.115 & 52.13 & 73.17 & 227.017 & 51.37 & 70.76 & 48.420 \\
\rowcolor[HTML]{EFEFEF} 
\midrule

& \textbf{\method}& 70.68 & 87.31 & 0.416 & 80.90 & 93.23 & 0.378 & 61.10 & 77.93 & 0.403 & 53.18 & 73.84 & 0.567 & 52.72 & 70.92 & 0.536 \\

& (vs.ColQwen2.5) & \sblue{102.65\%} & \sblue{101.99\%} & \sred{-99.22\%} & \sblue{98.72\%} & \sblue{99.96\%} & \sred{-99.65\%} & \sblue{102.84\%} & \sblue{101.13\%} & \sred{-98.98\%} & \sblue{102.02\%} & \sblue{100.92\%} & \sred{-99.75\%} & \sblue{102.62\%} & \sblue{100.23\%} & \sred{-98.89\%} \\ 
\bottomrule
\end{tabular}
}
\caption{Efficiency–accuracy comparison with various single-vector and multi-vector models for visual document retrieval on \benchmark performed within each data split. We report Recall@\{1,3\} and per-query FLOPs (billions) for \method, compared with both single-vector and multi-vector models. The relative performance (\%) for Recall is highlighted in \blue{blue}, and the FLOPs is highlighted in \red{red}. \textcolor{gray}{\faFileTextO} indicates a textual embedding model and \textcolor{gray}{\faPictureO} indicates a visual embedding model.}
\label{tab:main-split-vimdoc}
\end{table*}

\begin{table*}[t]
\centering

\setlength\tabcolsep{1.7pt} 
\setlength{\extrarowheight}{2pt}
\scalebox{0.67}{
\begin{tabular}{cl|ccc|ccc}
\toprule

& & \multicolumn{6}{c}{\textbf{OpenDocVQA}} \\
\cmidrule{3-8}
& \textbf{Data Split} & \multicolumn{3}{c|}{\textbf{SlideVQA}} & \multicolumn{3}{c}{\textbf{DUDE}} \\

\midrule
& \textbf{Model} & \textbf{R@1} & \textbf{R@3} & \textbf{FLOPs} & \textbf{R@1} & \textbf{R@3} & \textbf{FLOPs} \\
\midrule

\multirow{5}{*}{\rotatebox[origin=c]{90}{Single-Vector}} & BGE-M3 (dense) \textcolor{gray}{\textsuperscript{\faFileTextO}} & 41.25 & 56.71 & 0.107 & 37.35 & 50.15 & 0.057 \\
&  NV-Embed-V2 \textcolor{gray}{\textsuperscript{\faFileTextO}} & 50.66 & 69.34 & 0.429 & 44.41 & 60.74 & 0.229 \\

& VisRAG \textcolor{gray}{\textsuperscript{\faPictureO}} & 57.50 & 75.26 & 0.241 & 44.71 & 59.66 & 0.129 \\

& GME \textcolor{gray}{\textsuperscript{\faPictureO}} & 59.67 & 78.46 & 0.161 & 49.83 & 66.75 & 0.086 \\

& DSE \textcolor{gray}{\textsuperscript{\faPictureO}} & 64.93 & 81.12 & 0.161 & 54.59 & 70.72 & 0.086 \\ 
\midrule

\rowcolor[HTML]{EFEFEF} 
& \textbf{\method (only Stage 1)}  & 64.34 & 80.79 & 0.089 & 55.49 & 72.33 & 0.058 \\
& (vs. DSE) & \sblue{99.09\%} & \sblue{99.59\%} & \sred{-45.00\%} & \sblue{101.66\%} & \sblue{102.28\%} & \sred{-31.98\%} \\
\midrule

\multirow{4}{*}{\rotatebox[origin=c]{90}{Multi-Vector}} &
BGE-M3 (multi) \textcolor{gray}{\textsuperscript{\faFileTextO}} & 46.71 & 63.36 & 221.422 & 50.05 & 65.47 & 420.909 \\

& Colpali \textcolor{gray}{\textsuperscript{\faPictureO}} & 71.45 & 84.91 & 479.763 & 63.56 & 80.19 & 193.785 \\
& ColQwen2 \textcolor{gray}{\textsuperscript{\faPictureO}} & 73.55 & 88.33 & 350.585 & 70.92 & 84.69 & 138.214 \\
& ColQwen2.5 \textcolor{gray}{\textsuperscript{\faPictureO}} & 73.46 & 89.32 & 350.585 & 72.77 & 83.08 & 138.214 \\
\midrule
\rowcolor[HTML]{EFEFEF} 

& \textbf{\method} & 73.27 & 87.81 & 0.544 & 70.55 & 81.84 & 0.36 \\
& (vs.ColQwen2.5) & \sblue{99.73\%} & \sblue{98.31\%} & \sred{-99.61\%} & \sblue{96.95\%} & \sblue{98.50\%} & \sred{-99.90\%} \\
\bottomrule
\end{tabular}
}
\caption{Efficiency–accuracy comparison with various single-vector and multi-vector models for visual document retrieval on OpenDocVQA performed within each data split. We report Recall@\{1,3\} and per-query FLOPs (billions) for \method, compared with both single-vector and multi-vector models. The relative performance (\%) for Recall is highlighted in \blue{blue}, and the FLOPs is highlighted in \red{red}. \textcolor{gray}{\faFileTextO} indicates a textual embedding model and \textcolor{gray}{\faPictureO} indicates a visual embedding model. }
\label{tab:main-split-opendoc}
\end{table*}

\end{document}